\documentclass[12pt,a4paper]{article}

\usepackage[margin=1in]{geometry} 
\usepackage{float}
\usepackage{appendix,soul}
\usepackage{natbib}
\usepackage{graphicx}
\usepackage{pdflscape}
\usepackage{tabularx,booktabs,array,dcolumn,threeparttable}
\usepackage{amsmath,amsthm,amssymb,bbold,amsfonts,mathtools,marvosym}
\usepackage{eurosym}
\usepackage{caption,subcaption}
\usepackage{rotating}
\usepackage{ragged2e}
\usepackage{setspace}
\usepackage{authblk}
\usepackage[pdftex,
            pdfauthor={Ignacio Belloc, José Alberto Molina},
            pdftitle={Households with insufficient liquid assets: Consumption responses to income changes},
            pdfsubject={D12; D14; D15; E21; G51},
            pdfkeywords={hand-to-mouth; savings; marginal propensity to consume; HFCS data},
            hyperfootnotes=false,bookmarksopen=true]{hyperref} 
\hypersetup{colorlinks=true,allcolors=blue}

\title{\textbf{Households with insufficient liquid assets: \\ Consumption responses to income changes}\thanks{
Belloc: Department of Economic Analysis, University of Zaragoza; \href{mailto:ibelloc@unizar.es}{ibelloc@unizar.es}. Molina: Department of Economic Analysis, University of Zaragoza; \href{mailto:jamolina@unizar.es}{jamolina@unizar.es}.
\newline \textit{Funding}: This paper has benefitted from funding from the Spanish Ministry of Science, Innovation and Universities [PID2024-156465NB-I00]. 
\newline \textit{Acknowledgements}: This paper was partially written while I. Belloc and J. A. Molina were visiting the Department of Econometrics \& Operations Research at Tilburg University (The Netherlands) and the Department of Economics at Boston College (US), to which they would like to express their thanks for the hospitality and facilities provided. We are also grateful for the helpful comments provided by participants in several conferences.
\newline \textit{Declaration of Competing Interest}: None.
}}

\author[1]{Ignacio Belloc}

\author[1,2]{José Alberto Molina}

\affil[1]{\small IEDIS and University of Zaragoza, Zaragoza, Spain}
\affil[2]{\small IZA, LISER, Luxembourg}

\date{\vspace{-0.5cm} \normalsize\today}

\begin{document}

%%% Title
\maketitle

%%% Abstract 
\begin{abstract}

The fraction of households living with insufficient liquid assets is important to understand consumption responses to income changes. Using harmonized data for 23 European countries over 2010--2023 from the Household Finance and Consumption Survey, we investigate the consumption responses to income changes of hand-to-mouth (HtM) households, grouped into non-HtM, poor HtM and wealthy HtM. Our findings indicate significant variability across countries in the shares of HtM households, with the majority in all countries being wealthy households with sizeable wealth in housing and other real estates. By examining the marginal propensity to consume (MPC) to hypothetical shocks across these households, we find that poor HtM households exhibit the highest MPC, whereas wealthy HtM households display a negative association with the MPC. The relationship for poor HtM households seems to be driven by unobserved factors, whereas the relationship with wealthy HtM is negative and significant even controlling for unobserved preference heterogeneity and households who switch their status. These consumption responses align with life-cycle models with liquidity constraints and precautionary savings.

\vspace{0.5em}

\vspace{0.5em}
\end{abstract}
\pagebreak

%%%%%%%%%%%%%%%%%%%%%%%%%%%%%%%%%%%%%%%%
%%% Introduction
%%%%%%%%%%%%%%%%%%%%%%%%%%%%%%%%%%%%%%%%

\section{Introduction}

Households allocate their current income between consumption and savings taking into account the persistence of current changes and the uncertainty about their future income flows. As a result, forward looking consumers adjust their consumption only in response to unexpected and permanent income shocks, against expected and transitory ones. From a theoretical perspective, these implications are formalized under the so-called Life-Cycle and Permanent Income models \citep{ModiglianiBrumberg1954, Friedman1957}. However, contrary to these economic theories on consumption behavior, anticipated and temporary changes to income components are not smoothed and also trigger meaningful responses in consumption (e.g. see \citealp{Blundell2008, Blundell2016, Jappelli2010, Crawley2020, Fagereng2021, Wu2021, Commault2022, Bryukhanov2024}). 

Households' wealth is composed of a wide variety of (safe and risky) assets that are stored in either financial markets or their homes, with possibly very different degrees of liquidity, returns and adjustment costs. One of the most prominent determinants in explaining the issue of excessive sensitivity of consumption to current income is liquidity constraints \citep{Attanasio2010, Kaplan2014, Fuchs2016, Fisher2020}. Specifically, the lack of liquid assets as a mechanism to self-insure against fluctuations of income through financial assets prevents households from responding to income risks and completely smooth consumption over the life-cycle.

This household heterogeneity in the consumption--income dynamics due to differences in assets holdings, which implies that marginal propensities to consume (MPCs) out of income changes are decreasing in liquid wealth, has stimulated research aiming to identify which consumers are more likely to be liquidity constrained and thus overrespond to changes in the household economic environment. However, any definition of liquidity constraints depends on household characteristics typically not observed in data sources (i.e., preference heterogeneity across households) that could drive the excess sensitivity of consumption and thus contaminate results, such as impatience and risk attitudes \citep{Aguiar2025, Jappelli2025}. As a result, it is difficult to empirically measure the degree of household-level liquidity constraints and different definitions have been proposed. 

For instance, \citet{Kaplan2014} (henceforth, KVW), use the Survey of Consumer Finances (SCF) and show that the fraction of households that are unable to save and have insufficient amounts of liquid wealth is around 31\% in the US. These households have been labeled as hand-to-mouth (HtM) households. Furthermore, they show that most of these households are rich because they hold positive, and sometimes substantial, illiquid wealth. They incorporate these households into the HtM pool and call them ``the wealthy HtM''. Although traditionally overlooked in the literature because of their sizeable wealth position due to their large illiquid asset holdings, these consumers are also important in explaining heterogeneous consumption responses, since they are still affected by liquidity constraints, and behave like HtM households because they hold low levels of liquid assets.

Other related approaches that analyze binding financial constraints include the fraction of liquid assets to household income \citep{Zeldes1989, Johnson2010, Campbell2019, Sokolova2023, Jappelli2025}, the position across household wealth and income deciles \citep{Johnson2006, Parker2013, Christelis2020, Fisher2020, Fagereng2021, Kim2025, Belloc2026}, the difficulty in obtaining loans for finance purchases \citep{Jappelli1990, Cox1993, Paiella2017, Toussaint2021, Sala2024, Castaldo2025}, and the ratio between debt payments relative to household income \citep{Johnson2010, Wu2021, DuCaju2023}.\footnote{Table A.3 in \citet{Sokolova2023} presents different definitions used for binding liquidity constraints.}

Recent research that updated the KVW estimates supports these figures using the modern Panel Study of Income Dynamics in the US \citep{Aguiar2025, Belloc2025}, which collects detailed information on wealth and consumption at a biennial frequency since 1999. Beyond the US, a limited set of estimates appears, with recent research indicating that this fraction ranges from 23\% in Belgium \citep{Cherchye2024} to 27\% in Portugal \citep{Duarte2023}, with Spain falling between these two extremes \citep{Cutanda2025}. Although \citet{Kaplan2014} primarily focus on the US economy, they also report the shares of HtM across Australia, Canada, the United Kingdom and a limited number of large European countries in 2010, namely Germany, France, Italy, and Spain, with figures indicating that less than 20\% of households in Spain and more than 30\% of households in Germany live hand-to-mouth, most of them being wealthy HtM households. Extending their analysis to 2014, \citet{Slacalek2020} and \citet{Almgren2022} obtain distinct averages, suggesting some variance over time in this measure within countries.

In other geographical contexts, \citet{Park2017} and \citet{Song2020} find shares of HtM households above 30\% in South Korea, whereas \citet{Hara2016} obtain that the share of HtM is about 13\% in Japan.\footnote{\citet{Song2020} did not use the same measure to define HtM households.} Although limited research attention has been devoted to developing countries due to data availability complications and very data-demanding requirements imposed by the empirical strategy of KVW, \citet{Cui2017} obtain that on average 17\% of households are HtM in China, whereas \citet{Gupta2025} use a different approach based on imputation techniques for household income and obtain that about 17--32\% are HtM among Indian households. Similarly to developed economies, the vast majority of HtM households consists of wealthy HtM.
\clearpage

Within this framework, the purpose of this paper is to investigate the consumption responses to income changes of HtM households using harmonized cross-country data in Europe from 2010--2023. These households with insufficient liquid assets are distinguished among poor and wealthy HtM through a survey question that asks what percentage of a transitory, unexpected, income change would hypothetically be transmitted into either spending or saving choices. This new evidence of heterogeneous MPCs across different non-HtM, poor and wealthy HtM households, rather than average MPCs for all households, allows us to verify the predictions of standard consumption models. Related works have shown that holdings of liquid wealth correlate with MPCs \citep{Jappelli2014, Jappelli2020, Christelis2019, Christelis2021, Fuster2021}, but low holdings of liquid assets include both those who have low net worth as well as those who are wealthy HtM. The representative panel of the population constitutes a key advantage over previous research \citep{Bunn2018, Drescher2020, Christelis2019, Christelis2021, Fuster2021, Albacete2025, Jappelli2026} and allows to identify households who switch their HtM status by controlling for unobserved preference heterogeneity across households. 

The remainder of the paper is organized as follows. Section~\ref{Sec::Data} describes the data we use in our analysis and provides descriptive statistics of the shares of HtM households in Europe. This includes shares by country and recent trends over time, alongside analyses of the characteristics of HtM households. Section~\ref{Sec::Methodology} outlines our econometric strategy, whereas Section~\ref{Sec::MPC} empirically analyzes the MPC using regression analysis across these household groups. Finally, Section~\ref{Sec::Conclusions} concludes the paper. The appendices contain additional results.

%%%%%%%%%%%%%%%%%%%%%%%%%%%%%%%%%%%%%%%%
%%%	Data
%%%%%%%%%%%%%%%%%%%%%%%%%%%%%%%%%%%%%%%%

\section{Data and variables}\label{Sec::Data}

\subsection{The HFCS dataset}

Our empirical analysis is based on micro-data from the Household Finance and Consumption Survey (henceforth, HFCS), a harmonized cross-country survey coordinated by the European Central Bank, jointly conducted with all the national central banks of the Eurosystem and EU countries that have not yet adopted the euro, as well as various national statistical institutes. The HFCS is a repeated cross-sectional survey of a representative sample of European households, in general and at country level, that has been conducted in five waves: 2010, 2014, 2017, 2021, and 2023.\footnote{The fieldwork period of the survey does not match across countries, both in terms of length and time period, but those years are the most common reference period for the data in each wave, except wave 2023 which refers to 2022 year.} For certain countries, it also includes a panel component for those five years. The main purpose of the HFCS is to provide insights into the distribution of household net wealth and its components in the euro area, so it is particularly suitable to carry out our analysis.\footnote{The HFCS is the only database that currently provides the most comprehensive information on income, assets and liabilities for all the population in Europe, compared to alternative datasets such as the European Union Statistics on Income and Living Conditions (EU-SILC), which focuses on income, or the Survey of Health, Ageing and Retirement in Europe (SHARE), which targets respondents aged 50 and above.} 

Our study uses data from all the years of the HFCS, the most recent data at the time of writing this article, and we combine the HFCS into a single pooled cross-section. The HFCS contains information on a sample of more than 410,000 households (68,627 were surveyed in 2010, 84,611 in 2014, 91,242 in 2017, 83,162 in 2021, and 84,755 in 2023) across all the euro area countries, as well as Czech Republic, Hungary and Poland, with resulting sample sizes varying in each country. For consistency, given that wealth components are collected at the household level in the data, as is frequent in household surveys \citep{Belloc2026}, we construct all of the variables at the household level. Hence, for individual-level demographic variables, we attribute to each household variables associated with the household head. 

\subsection{Sample requirements}

Turning to the sample selection of our empirical analysis, we start with a sample of household heads and mimic all of the sample restrictions made by KVW (we refer the reader to their paper for a detailed discussion of the motivation behind those restrictions). In so doing, we select household heads aged 22--79, both included, i.e., who make active consumption and saving decisions. This removes 33,962 households. We do not select our sample based on the working status of the household head or spouse (if any) to minimize sample selection biases and only exclude households who derive all their income from self-employment (i.e., about 6,988 households) to eliminate the impact of any unusual observations, as it is likely that both assets and income from these are not well measured.\footnote{In our sample, there are no households with negative or missing income, conditional on the above restrictions.} Finally, we retain households with complete information on key variables such as demographics, income, work status, and wealth (overall, these restrictions discard around 1.3\% of observations/households). The HFCS imputes data for missing values for some variables and uses multiple imputation techniques to minimize missing observations. This consists of 5 different imputed values (`implicates') for selected variables. The restrictions should apply consistently to all the implicates per household.

The restrictions leave us with a final sample corresponding to 349,908 household-year observations across 5 waves (2010, 2014, 2017, 2021, and 2023) from 23 European countries, namely: Austria (AT), Belgium (BE), Cyprus (CY), Czech Republic (CZ), Germany (DE), Estonia (EE), Spain (ES), Finland (FI), France (FR), Greece (GR), Hungary (HU), Croatia (HR), Italy (IT), Ireland (IE), Lithuania (LT), Luxembourg (LU), Latvia (LV), Malta (MT), Netherlands (NL), Poland (PL), Portugal (PT), Slovenia (SI), and Slovakia (SK).\footnote{Note that the survey has a panel component, meaning that some of those households are followed over time.} Of these, 12,120 observations correspond to Austria, 9,916 to Belgium, 5,915 to Cyprus, 5,581 to Czech Republic, 18,655 to Germany, 8,986 to Estonia, 26,236 to Spain, 41,917 to Finland, 56,477 to France, 13,918 to Greece, 23,036 to Hungary, 3,722 to Croatia, 5,072 to Ireland, 34,100 to Italy, 4,587 to Lithuania, 9,552 to Luxembourg, 4,555 to Latvia, 4,477 to Malta, 9,611 to Netherlands, 8,615 to Poland, 24,754 to Portugal, 8,122 to Slovenia, and 9,984 to Slovakia.

\subsection{Variable construction}

The HFCS collects detailed information on household portfolios and debts. Specifically, the household head or the person most knowledgeable about the family's finances is asked about ownership of many types of assets and liabilities in various categories (i.e., the participation rate), and about the amounts of wealth held and owed in each category (the intensive margin). Assets include: cash, savings and deposits, stocks, bonds, cash value in life insurance, home value, and other investments such as vehicles, business assets, valuables, and other real estate.\footnote{Valuables are defined as valuable jewelry, antiques or art.} On the other hand, liabilities are divided into mortgage debt and non-mortgage debt, where the former consists of mortgages on the household main residence and mortgages on other real estate properties, and the latter comprises credit line and overdraft debt, credit card debt and other non-mortgage loans.

From the information provided in the survey, we classify assets into liquid and illiquid wealth components. Our division of assets by liquidity largely follows the definitions of previous research using similar data \citep{Kaplan2014, Slacalek2020, Cherchye2024, Aguiar2025, Cutanda2025}. Specifically, we define liquid wealth as the sum of cash, balances in sight and savings accounts (i.e., current accounts), mutual fund holdings directly held, shares in publicly traded companies, and corporate or government bond holdings, whereas we measure illiquid wealth as the sum of the value of the household's main residence and other real estate properties, plus the value of any occupational and voluntary pension plans, the cash value of life insurance policies, and certificates of deposit.\footnote{We do not include other nonfinancial wealth, such as antiques, artwork, and jewels, or the value of self-employment businesses in this definition, as in \citet{Kaplan2014} or \citet{Cherchye2024}.} In terms of debt, we consider the balance on credit cards and the balances on any credit lines or bank overdrafts as liquid debts in order to calculate net liquid wealth, and debts related to real estate acquisitions (i.e., mortgages) are considered as illiquid debt to obtain net illiquid wealth.\footnote{Financial debts, such as overdrafts or credits tied to household consumption, are more easily liquidated relative to real debt products.} 

We define a household net worth as the sum of its liquid and illiquid assets net of debts. These components of net worth are the key variables in order to properly define and stratify households into poor HtM, wealthy HtM and non-HtM, in terms of either net liquid and illiquid wealth or net worth according to the definitions proposed in \citet{Kaplan2014} and \citet{Zeldes1989}, respectively (see Subsection~\ref{Subsec::HtM} below for a discussion on this issue).\footnote{In order to decrease non-response rates, the HFCS uses stochastic multiple imputation techniques for all missing observations in households' income, consumption and wealth. We estimate all our results using these imputed values together, adjusting coefficients and standard errors accordingly throughout the paper (see \citealp{Trivin2022, DuCaju2023, Sala2024}).} 

Furthermore, in order to categorize a given household as either poor or wealthy HtM, we also need information on household income. Our measure of income in the survey is defined as the sum of all income earned by household members over the last 12 months preceding the survey or the calendar year prior to the survey year. It includes both labor and non-labor income sources such as gross income from wages, salaries, self-employment, and pension, unemployment benefits, and regular transfers such as child support alimony and other public transfers, consistent with \citet{Kaplan2014}, \citet{Cherchye2024} and \citet{Aguiar2025}. We exclude interest, dividends, and other capital income because they are realized more infrequently. All these monetary amounts, i.e. household income and wealth, are deflated and expressed in 2021 constant euros using the Harmonised Index of Consumer Prices (HICP) from the Eurostat data (the annual average index, all items HICP).\footnote{Available at \href{https://ec.europa.eu/eurostat/databrowser/view/prc\_hicp\_aind/default/table?lang=en}{https://ec.europa.eu/eurostat/databrowser/view/prc\_hicp\_aind/default/table?lang=en}, accessed January 2026.}

Finally, the HFCS allows us to define a range of other household and individual demographics that may be important in explaining HtM and have been shown to be important determinants of consumption behavior. These variables include the age of the head of the household, a dummy variable indicating whether the household head is employed, variables denoting the maximum educational level achieved by the household head, primary housing tenure status, the marital status through a dummy variable denoting whether the household head is married, the number of individuals in the family unit, and the number of children under the age of 16. Education is categorized into three groups based on the maximum level of educational attainment achieved by the household head on the basis of the 1997 International Standard Classification of Education (ISCED-97) scale: primary or no education comprises heads who do not have secondary education, secondary education includes those who completed secondary education but did not graduate, and tertiary education refers to those with tertiary education. Finally, we control for the primary housing tenure status and define three dummy variables indicating households who own their primary residence without mortgage or rental payments (i.e., outright owners), households who own part of the residence and have a mortgage (homeowners with a mortgage), and households who are renters.

\subsection{Identifying HtM households in Europe}\label{Subsec::HtM}

We identify HtM households following the criteria of \citet{Kaplan2014} and split households between HtM and non-HtM based on their holdings of net liquid and illiquid wealth. Specifically, households whose balances of liquid wealth are positive but equal to or less than half their earnings per pay period are classified as HtM.\footnote{As in KVW, we set the pay frequency to two weeks.} In addition, we also incorporate the household's credit limit, and count as HtM any household whose balances of liquid wealth are negative and for whom the sum of their liquid wealth and credit limit (assumed at 0.185 times of annual earnings \citep{Kaplan2014, Aguiar2025}) is less than half their earnings per pay period. Next, to distinguish between poor and wealthy HtM among households who do not hold liquid assets (i.e., conditional on being HtM), we rely on their holdings of net illiquid wealth, if any. Specifically, households who hold \textit{any} amount of illiquid wealth are labeled as wealthy HtM, whereas those who hold negative or zero amounts of illiquid wealth are labeled as poor HtM. 

\begin{figure}[H]
\centering

\caption{Shares of total, wealthy, and poor HtM households, HFCS 2010--2023}
\label{fig::HtM_shares}

\begin{threeparttable}
    \includegraphics[width=1\textwidth]{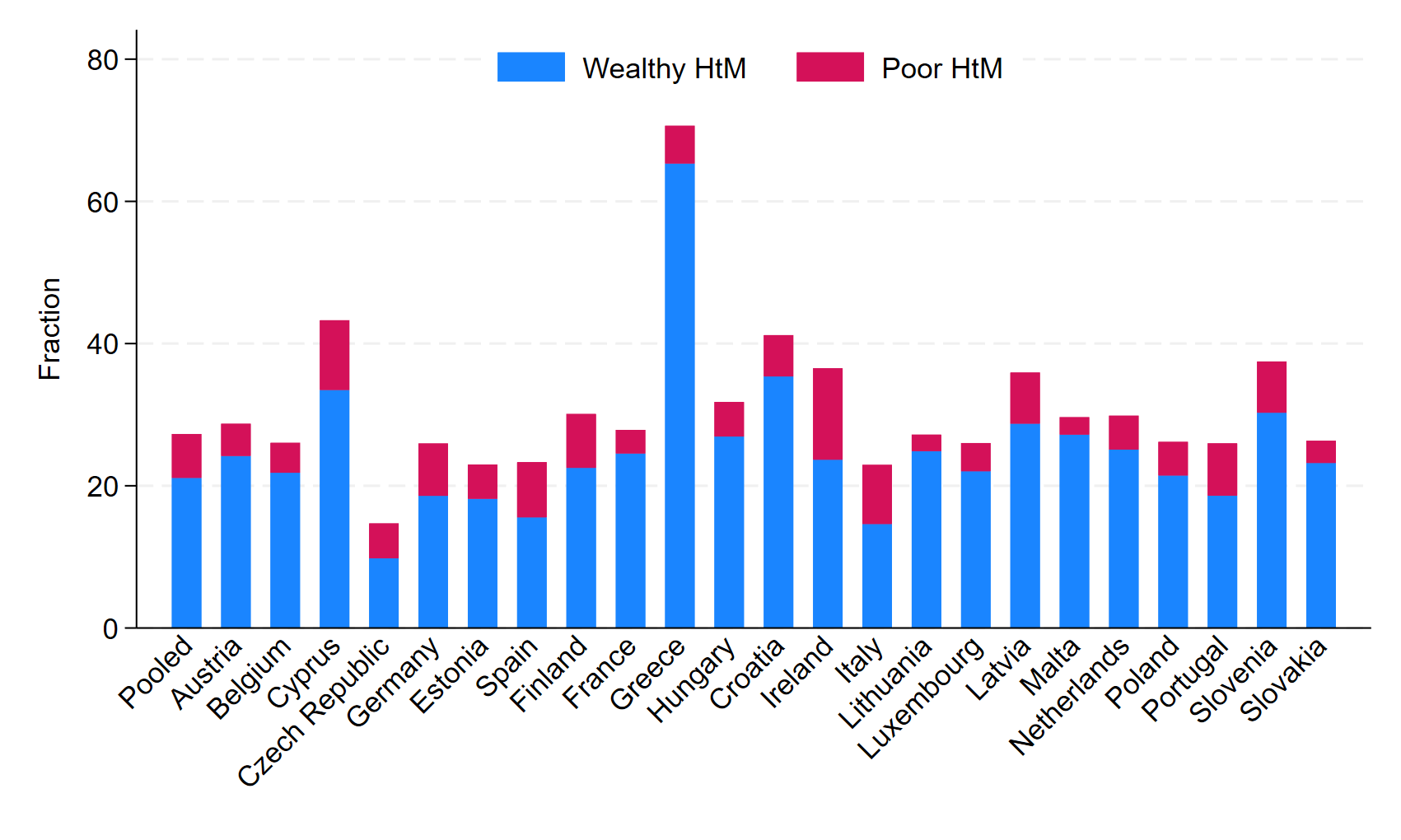}
    \begin{tablenotes}
        \footnotesize
        \item \textit{Notes:} The figure shows percentages across HtM households for the different countries. Percentages over \textit{total} households. Data come from the Household Finance and Consumption Survey, 2010--2023. Figures refer to sample averages, calculated using population weights and bootstrapped replicated weights. 
    \end{tablenotes}
\end{threeparttable}

\end{figure}

We present summary statistics on HtM households across countries and over time, given the extended time span covered by the dataset. Figure~\ref{fig::HtM_shares} shows the fraction of HtM households in our sample of European countries, distinguishing between wealthy, poor, and total HtM households, using the pooled cross-sectional HFCS dataset over the 2010--2023 period. Across our entire sample, we label 27.3\% of households as HtM. The fraction of wealthy HtM is, on average, larger than the share of poor HtM, with 21.1\% denoted as wealthy HtM and the remaining 6.2\% as poor HtM. 

Across all countries, HtM households constitute a quantitatively important share of the population, with HtM shares ranging from roughly 14\% in Czechia to nearly 71\% of households in Greece. Apart from these countries, other countries such as Estonia, Spain and Italy have low levels under 25\%, and others have values higher than 40\%, such as Cyprus and Croatia. Moreover, in every country the majority (i.e., well over 65\%) of HtM households are wealthy rather than poor, indicating that these households hold substantial wealth in illiquid assets, a common finding in the existing literature. In particular, the share of wealthy HtM households among all HtM households varies between 64\% in Ireland and Italy and 90\% in Greece, Luxembourg, and Malta.

As expected, the variation across countries is significant. Among Western European countries, the fraction of HtM households ranges from 25.9\% in Germany to 36.5\% in Ireland. In Eastern European countries, HtM shares lie between 13.9\% in Czech Republic and 43.3\% in Cyprus. Southern European countries display the greatest cross-sectional variance, largely driven by Greece and Cyprus, with HtM shares oscillating between 22.9\% in Spain and Italy and 70.6\% in Greece.

Alternatively, Table~\ref{tab::HtM_time} documents the evolution of the share of HtM households over time for all countries in the sample.\footnote{The patterns in some subsets of countries, such as Belgium, Spain, and Portugal, match with recent research at the country level using part of the HFCS \citep{Kaplan2014, Slacalek2020, Duarte2023, Cherchye2024, Cutanda2025}.} Overall, the share of HtM households experiences a decline in our sample over 2010--2023, from 28.5\% in 2010 to 27.3\% in 2023, especially concentrated during the last two waves of the survey and due to the lower presence of wealthy HtM households. In contrast, the share of poor HtM households is relatively stable over time. Most countries exhibit a decline in the prevalence of HtM households over the sample period, including Austria, Belgium, Cyprus, Czech Republic, Germany, Estonia, Spain, Finland, Hungary, Croatia, Italy, Lithuania, Luxembourg, Latvia, Malta, the Netherlands, Slovenia, and Slovakia. By contrast, other countries, such as France, Greece, and Portugal, display an increase in these shares over time. Finally, the fraction of HtM households has stayed stable in Poland throughout the observation period.

\begin{sidewaystable}[htp]
\begin{center}
\caption{Trends in shares of HtM households}\label{tab::HtM_time}
\resizebox{0.95\textwidth}{!}{
\begin{tabular}{lccccccccccccccccccccccc}
\toprule
 & \multicolumn{3}{c}{2010} & & \multicolumn{3}{c}{2014} & & \multicolumn{3}{c}{2017} & & \multicolumn{3}{c}{2021} & & \multicolumn{3}{c}{2023} & & \multicolumn{3}{c}{2010--2023} \\
\cmidrule(lr){2-4} \cmidrule(lr){6-8} \cmidrule(lr){10-12} \cmidrule(lr){14-16} \cmidrule(lr){18-20} \cmidrule(lr){22-24}
 & Wealthy & Poor & Total & & Wealthy & Poor & Total & & Wealthy & Poor & Total & & Wealthy & Poor & Total & & Wealthy & Poor & Total  & & Wealthy & Poor & Total \\ \midrule
Austria  & 29.7 & 4.9 & 34.5 & & 26.1 & 5.2 & 31.3 & & 21.9 & 3.9 & 25.8 & & 22.4 & 4 & 26.4 & & 21.3 & 4.7 & 26 & & 24.2 & 4.5 & 28.7  \\ 
Belgium  & 22.2 & 4 & 26.2 & & 21.5 & 6.4 & 27.9 & & 19.7 & 5.3 & 25 & & 22.7 & 5.2 & 27.9 & & 23 & 0.4 & 23.5 & & 21.8 & 4.2 & 26 \\ 
Cyprus  & 37.7 & 5.1 & 42.8 & & 33.3 & 10.6 & 43.9 & & 37.8 & 11.5 & 49.4 & & 34.3 & 11.6 & 45.9 & & 25.8 & 10.1 & 35.9 & & 33.5 & 9.8 & 43.3 \\ 
Czech Republic & --- & --- & --- & & --- & --- & --- & & --- & --- & --- & & 9.9 & 4.3 & 14.2 & & 9.7 & 3.9 & 13.6 & & 9.8 & 4.1 & 13.9 \\
Germany  & 22.7 & 7.3 & 30 & & 20.9 & 8 & 28.9 & & 19.3 & 8.1 & 27.5 & & 15.3 & 6.1 & 21.4 & & 14.8 & 7.2 & 22 & & 18.6 & 7.4 & 25.9 \\ 
Estonia  & --- & --- & --- & & 21.1 & 7.3 & 28.4 & & 18 & 6.3 & 24.3 & & 16.8 & 3.5 & 20.2 & & 17 & 2.4 & 19.4 & & 18.2 & 4.8 & 23 \\ 
Spain  & 18 & 5.9 & 23.8 & & 17 & 7.6 & 24.6 & & 14.7 & 7.8 & 22.5 & & 15 & 9.5 & 24.5 & & 13.4 & 8 & 21.4 & & 15.6 & 7.8 & 23.3 \\ 
Finland  & 27.5 & 8.9 & 36.3 & & 21 & 6.8 & 27.8 & & 21.6 & 7.2 & 28.8 & & 21.9 & 7.3 & 29.2 & & 23 & 8.3 & 31.3 & & 22.5 & 7.6 & 30.1 \\ 
France  & 23.4 & 3.1 & 26.5 & & 25.7 & 3.1 & 28.8 & & 25.8 & 3.9 & 29.7 & & 24.4 & 3.3 & 27.7 & & 23.5 & 3.2 & 26.7 & & 24.5 & 3.3 & 27.9 \\ 
Greece  & 26.9 & 9.5 & 36.5 & & 68.1 & 10.3 & 78.4 & & 76.8 & 5.2 & 82 & & 74.4 & 1.3 & 75.8 & & 77.6 & 0.7 & 78.3 & & 65.3 & 5.3 & 70.6 \\ 
Hungary  & --- & --- & --- & & 29.7 & 5.3 & 35 & & 33 & 6.5 & 39.5 & & 20.8 & 3.2 & 24 & & 24.1 & 4.4 & 28.6 & & 26.9 & 4.9 & 31.8 \\ 
Croatia  & --- & --- & --- & & --- & --- & --- & & 44.4 & 5.9 & 50.3 & & 35.1 & 7.7 & 42.8 & & 26.4 & 3.8 & 30.2 & & 35.4 & 5.8 & 41.2 \\ 
Ireland & --- & --- & --- & & 23.7 & 12.9 & 36.5 & & --- & --- & --- & & --- & --- & --- & & --- & --- & --- & & 23.7 & 12.9 & 36.5 \\
Italy  & 17.8 & 8.5 & 26.2 & & 16.3 & 10.5 & 26.8 & & 15 & 9.6 & 24.6 & & 13.1 & 6.7 & 19.8 & & 11.1 & 6.3 & 17.4 & & 14.6 & 8.3 & 22.9 \\ 
Lithuania  & --- & --- & --- & & --- & --- & --- & & 26.7 & 3 & 29.7 & & 26.5 & 1.1 & 27.6 & & 21.9 & 2.7 & 24.6 & & 24.9 & 2.3 & 27.2 \\ 
Luxembourg  & 19 & 5.7 & 24.7 & & 23.7 & 5.5 & 29.3 & & 23.7 & 4.4 & 28.1 & & 25.3 & 2.7 & 28 & & 18.1 & 2.4 & 20.5 & & 22.1 & 4 & 26 \\ 
Latvia  & --- & --- & --- & & 39.8 & 11.3 & 51.1 & & 36.1 & 9.7 & 45.8 & & 20.3 & 3.5 & 23.8 & & 18.7 & 4.1 & 22.8 & & 28.7 & 7.2 & 35.9 \\ 
Malta & 34.9 & 3 & 37.9 & & 33.8 & 2 & 35.8 & & 36.8 & 1.4 & 38.3 & & 22.3 & 3.8 & 26.1 & & 14.8 & 2 & 16.7 & & 27.2 & 2.4 & 29.6 \\
Netherlands  & 34.7 & 4.8 & 39.5 & & 33.2 & 5 & 38.2 & & 17.5 & 5.5 & 22.9 & & 22.4 & 4.8 & 27.1 & & 18.4 & 3.8 & 22.2 & & 25.1 & 4.8 & 29.9 \\ 
Poland  & --- & --- & --- & & 20.7 & 5.6 & 26.4 & & 22.2 & 3.8 & 26 & & --- & --- & --- & & --- & --- & --- & & 21.4 & 4.7 & 26.2 \\ 
Portugal  & 18.4 & 5.7 & 24.1 & & 19.6 & 8.2 & 27.8 & & 19 & 8 & 27 & & 18.7 & 7.1 & 25.7 & & 17.5 & 7.8 & 25.3 & & 18.6 & 7.3 & 26 \\ 
Slovenia  & 35.4 &  6.5 & 41.8 & & 35 & 8.1 & 43.1 & & 26.6 & 8.4 & 35 & & 29.9 & 7.9 & 37.8 & & 24.6 & 5.2 & 29.8 & & 30.3 & 7.2 & 37.5 \\ 
Slovakia  & 22.7 & 2.3 & 25 & & 25.1 & 4.9 & 30 & & 26.7 & 3.4 & 30.1 & & 20.7 & 2.3 & 23.1 & & 20.7 & 2.7 & 23.4 & & 23.2 & 3.1 & 26.3 \\ 
& & & & & & & & & & & & & & & & & & & & & & & \\
\textbf{Pooled}  & \textbf{22.4} & \textbf{6.1} & \textbf{28.5} & & \textbf{23} & \textbf{7} & \textbf{30.1} & & \textbf{21.8} & \textbf{6.6} & \textbf{28.5} & & \textbf{19.7} & \textbf{5.6} & \textbf{25.3} & & \textbf{18.7} & \textbf{5.4} & \textbf{24.1} & & \textbf{21.1} & \textbf{6.2} & \textbf{27.3} \\ \bottomrule
\end{tabular} }
\begin{minipage}{0.95\textwidth}
\footnotesize
\textit{Notes}: Data come from the Household Finance and Consumption Survey, 2010--2023. Figures refer to sample averages, calculated using population weights and bootstrapped replicated weights. 
\end{minipage}
\end{center}
\end{sidewaystable}

\subsection{Descriptive statistics}

Table~\ref{tab::HtM_characteristics} presents summary statistics for five household groups: the full sample after applying the selection criteria, non-HtM households, HtM households, wealthy HtM households, and poor HtM households. We first compare HtM households with their non-HtM counterparts. On average, HtM households are approximately three years younger and display lower household income and labor earnings, consistent with \citet{Aguiar2025}. They are also characterized by negative net liquid wealth and substantially lower holdings of illiquid assets, resulting in lower overall net worth. In demographic terms, HtM households tend to be slightly larger, display a more balanced gender distribution among household heads and have lower levels of education, whereas they are less likely to be employed or married. Housing tenure also differs sharply across groups; while around 42\% of HtM households are renters, approximately 44\% of non-HtM households are outright owners.

Further insights emerge when comparing wealthy and poor HtM households. Specifically, heads of wealthy HtM households are, on average, about three years older and earn nearly twice as much in terms of both household income and labor earnings. Although wealthy HtM households hold substantial illiquid assets (about \euro156,311 in net illiquid wealth), their net liquid wealth remains negative and is even lower than that of poor HtM households. Notably, wealthy HtM households exhibit the lowest average in net liquid wealth among all groups, highlighting the liquidity constraints that also characterize this group despite their sizable illiquid asset position. Alongside other demographics, wealthy HtM households show the highest employment rates and larger family sizes, whereas poor HtM households have the lowest levels of educational attainment. As expected, poor HtM households tend to be renters (about 78.8\% state they are renters), whereas about 65\% of wealthy HtM households are home-owners (40.4\% are outright owners and 24.4\% mortgagors).

In Table~\ref{tab::HtM_debt} we look at additional characteristics among those groups of the population and analyze their level of indebtedness, the debt-to-income ratios, and the debt-payments-to-income ratio \citep{Johnson2010, Albuquerque2019, Toussaint2021, DuCaju2023, Cherchye2024, Sala2024, Bartscher2025, Belloc2025}, along with the presence of binding credit constraints. The debt-payments-to-income is the ratio between the household's total debt payments and disposable income. We also display the average ratio of total homeowners' mortgage balances and household income (i.e., the housing debt-to-income ratio). To control for binding financial constraints due to credit constraints, we consider whether the household experienced credit refusal within the last years or did not apply for credit within the last years as it expected to be turned down. (Note that we exclude 7,570 households who have missing values on debt payments or report zero amounts for family income to properly define the debt-payments-to-income and the debt-to-income ratios).\footnote{The survey question about the presence of credit constraints is not available in the HFCS for Italy in 2010 and 2014, for instance.}

The main difference across household groups arises from the composition and burden of debt, particularly in liquid liabilities unrelated to real estate. HtM households hold substantially larger amounts of liquid debt than non-HtM households and the population as a whole. In addition, HtM households also display higher debt payments related to income, as expected, and show a higher probability, more than double, of being credit constrained compared to their non-HtM counterparts. Within the HtM group, important heterogeneity also emerges. Wealthy HtM households hold the largest amounts of both mortgages and liquid debts. However, despite their higher indebtedness, they face a considerably lower debt burden relative to income and a substantially lower probability of being credit constrained than poor HtM households.

Finally, Table~\ref{tab::Transition} reports the transition probabilities across the HtM categories, allowing us to assess the persistence of household types by comparing households observed over two and three consecutive waves. Looking at the diagonal elements, we find the highest persistence among non-HtM households, as indicated by the first element of the main diagonal. In particular, households classified as non-HtM in a given year (Non-HtM$_t$, column) have an 86\% probability of remaining in that status in the following period and an 87\% probability two periods later. In contrast, households classified as wealthy or poor HtM in a given year exhibit only a  38.4--39.9\% probability of remaining in the same category in the following observation period. After two periods, persistence in HtM status declines. Overall, these findings regarding the persistence of household types are consistent with \citet{Aguiar2025}.\footnote{The period across HFCS interviews varies between 2-4 years. If we study average transition rates between HtM status across HFCS 2021-2023 waves, we find that the persistence across non-HtM, wealthy HtM and poor HtM is 88.6\%, 37.2\% and 41.2\%, respectively. That is, the persistence among poor HtM households is even larger.}

\begin{table}[H]
\begin{center}
\caption{Summary statistics among household groups}\label{tab::HtM_characteristics}
\resizebox{0.95\textwidth}{!}{
\begin{tabular}{lccccc}
\toprule
 & All &  Non-HtM & HtM & Wealthy HtM & Poor HtM \\ \midrule
\textit{Income and wealth} & & & & & \\
~~Income & 45,368.910 & 48,026.27 & 38,287.470 & 42,506.830 & 23,850.310 \\
~~Labor income (age 22--65) & 37,603.010 & 40,616.940 & 30,263.190 & 34,451.240 & 16,524.850 \\
~~Net liquid wealth & 17,855.040 & 28,948.660 & -11,707.600 & -13,304.960 & -6,242.147 \\
~~Net illiquid wealth & 206,038.700 & 238,325 & 120,001.300 & 156,310.600 & -4,220.173 \\
~~Net worth & 223,893.700 & 267,273.600 & 108,293.700 & 143,005.700 & -10,462.320 \\
& & & & & \\
\textit{Demographics} & & & & & \\
~~Male & 0.559 & 0.573 & 0.523 & 0.525 & 0.513 \\
~~Age & 51.949 & 52.807 & 49.664 & 50.301 & 47.483 \\
~~Primary education & 0.119 & 0.107 & 0.153 & 0.140 & 0.199 \\
~~Secondary education & 0.582 & 0.563 & 0.631 & 0.619 & 0.671 \\
~~Tertiary education & 0.299 & 0.330 & 0.216 & 0.241 & 0.130 \\
~~Employed & 0.601 & 0.604 & 0.593 & 0.623 & 0.490 \\
~~Married & 0.526 & 0.540 & 0.490 & 0.525 & 0.369 \\
~~Household size & 2.375 & 2.346 & 2.453 & 2.507 & 2.267 \\
~~Number of children & 0.412 & 0.381 & 0.497 & 0.494 & 0.505 \\
~~Outright owner & 0.403 & 0.436 & 0.313 & 0.404 & 0.002 \\
~~Mortgagor & 0.216 & 0.221 & 0.203 & 0.244 & 0.063 \\
~~Renter & 0.329 & 0.295 & 0.422 & 0.315 & 0.788 \\
\bottomrule
\end{tabular} }
\begin{minipage}{0.95\textwidth}
\footnotesize
\textit{Notes}: Data come from the Household Finance and Consumption Survey, 2010--2023. Figures refer to sample averages, calculated using population weights and bootstrapped replicated weights. All monetary figures are expressed in 2021 euros.
\end{minipage}
\end{center}
\end{table}

\begin{table}[H]
\begin{center}
\caption{Indebtedness among household groups}\label{tab::HtM_debt}
\resizebox{0.90\textwidth}{!}{
\begin{tabular}{lccccc}
\toprule
 & All & Non-HtM & HtM & Wealthy HtM & Poor HtM \\ \midrule
~~Mortgage debt & 32,100.850 & 33,237.780 & 29,080.680 & 31,972.320 & 18,472.990 \\
~~Liquid debt & 4,665.021 & 1,396.857 & 13,346.990 & 15,048.270 & 7,106.283 \\
~~Total debt-to-income ratio & 1.574 & 0.994 & 3.115 & 2.451 & 5.526 \\
~~Housing debt-to-income ratio & 1.385 & 0.962 & 2.509 & 1.865 & 4.853 \\
~~Debt-payments-to-income & 0.010 & 0.007 & 0.019 & 0.016 & 0.030 \\
~~Credit constraints & 0.076 & 0.057 & 0.128 & 0.108 & 0.202 \\
\bottomrule
\end{tabular} }
\begin{minipage}{0.90\textwidth}
\footnotesize
\textit{Notes}: Data come from the Household Finance and Consumption Survey, 2010--2023. Figures refer to sample averages, calculated using population weights and bootstrapped replicated weights. Debt figures, in levels, are expressed in 2021 euros.
\end{minipage}
\end{center}
\end{table}

\begin{table}[H]
\begin{center}
\caption{Transition rates for HtM households}\label{tab::Transition}
\resizebox{0.70\textwidth}{!}{
\begin{tabular}{lccc}
\toprule
 & Non-HtM$_{t}$ & Wealthy HtM$_{t}$ & Poor HtM$_{t}$ \\ \midrule
\textit{Panel A: 54,179 households} & & \\
Non-HtM$_{t+1}$ & 0.861 & 0.582 & 0.433 \\
Wealthy HtM$_{t+1}$ & 0.120 & 0.384 & 0.169 \\ 
Poor HtM$_{t+1}$ & 0.018 & 0.034 & 0.399 \\ \midrule
\textit{Panel B: 18,832 households} & & \\
Non-HtM$_{t+2}$ & 0.868 & 0.648 & 0.477 \\
Wealthy HtM$_{t+2}$ & 0.114 & 0.315 & 0.179 \\
Poor HtM$_{t+2}$ & 0.019 & 0.037 & 0.344 \\
\bottomrule
\end{tabular} }
\begin{minipage}{0.70\textwidth}
\footnotesize
\textit{Notes}: Data come from the Household Finance and Consumption Survey, 2010--2023.
\end{minipage}
\end{center}
\end{table}

\subsection{Robustness checks about the definition of HtM}

We run many robustness checks on our results for HtM households in Europe. For comparison, we also stratify HtM households in net worth based on \citet{Zeldes1989}, and consider that a household is HtM in net worth when their net worth is positive but equal to or less than half their earnings per pay period, or negative with absolute value greater than half their earnings per paid period plus the credit limit \citep{Kaplan2014, Aguiar2025}. This highlights the importance of distinguishing between liquid and illiquid asset positions. We also focus on a sample of 250,172 household heads aged 25-64 and use labor earnings as the reference measure for income \citep{Aguiar2025, Cutanda2025}. The percentages, shown in Table~\ref{Table::Additional_HtM} in Appendix~\ref{Appendix::A}, are significantly lower than those using the definition of KVW. Furthermore, we also change the payment frequency of income payments from two weeks to one month, which increases the fraction of HtM households. Symmetrically, shortening the paid period and setting it to one week decreases the fraction of HtM households. Finally, our main definition of HtM omits other real assets such as the value of self-employment business \citep{Slacalek2020} or other assets such as vehicles, jewelry, antiques or artwork, meaning that our baseline shares for wealthy HtM households outlined above may provide a lower bound. As a result, we explore an alternative specification of illiquid asset categories and incorporate those into the definition of net illiquid wealth, together with 6,869 households whose income is entirely from self-employment \citep{Kaplan2014}. This increases the fraction of wealthy HtM households, while the overall percentage of HtM households remains stable, as in KVW. All these results are included in Table~\ref{Table::Additional_HtM2}.

%%%%%%%%%%%%%%%%%%%%%%%%%%%%%%%%%%%%%%%%
%%%	Econometric strategy
%%%%%%%%%%%%%%%%%%%%%%%%%%%%%%%%%%%%%%%%

\section{Econometric strategy}\label{Sec::Methodology}

Since the 2017 survey wave, the HFCS collects information on the MPC per household. Specifically, the survey asks the following question: ``\textit{Imagine that you unexpectedly receive money from a lottery, equal to the amount of income your household receives in a month. What percentage would you spend over the next 12 months on goods and services, as opposed to any amount you would save for later or use to repay loans?}'', with the potential answers ranging from 0 (not spend anything) to 100 percent (spend the full windfall over the next 12 months), both included. This question does not target a specific category of consumption and refers to both durable and non-durable consumption spending. An additional question is available to measure the marginal propensity to save (MPS), analogous to the previous one (i.e., the sum of these two hypothetical choices must add up to 100 percent).\footnote{This question is nearly identical to a hypothetical question for spending responses available in the Italian Survey of Household Income and Wealth (SHIW) \citep{Jappelli2014, Jappelli2020} and does not refer to a fixed sum of money across households \citep{Crossley2021, Albuquerque2023}. A difference with the SHIW is that the question in the HFCS refers to an explicit time frame for the spending response. Similar data also exist for the Netherlands \citep{Christelis2019}.} 

This survey question directly elicits the MPC and captures the percentage by which households would raise their consumption upon receiving an unexpected, transitory and positive, income change within a year. A number of recent studies have used survey data and responses to scenarios involving changes to the households' economic environment to elicit the MPC (see, e.g., \citealp{Jappelli2014, Jappelli2020, Bunn2018, Christelis2019, Christelis2021, Fuster2021, Jappelli2026}) or labor supply responses \citep{Georgarakos2025}. 

In contrast to other approaches taken in the literature based on quasi-experimental events that mimic unexpected changes in household budgets (due, for example, to lottery wins, see \citealp{Fagereng2021, Golosov2024}) and semistructural methods based on a statistical decomposition of income processes and consumption \citep{Blundell2008, Ghosh2025}, this method uses a ``reported preference'' approach \citep{Fuster2021} and measures directly how households would modify their expenses in response to a scenario involving an unexpected windfall gain equal to their average monthly income.\footnote{\citet{Crawley2024} and \citet{Trivin2024} survey the advantages and underlying assumptions of this method to infer MPCs, whereas \citet{Stantcheva2024} and \citet{Ueda2025} compare self-reported and actual spending behavior.}

We go beyond the previous descriptive evidence and further analyze this unique survey question available in the HFCS using formal regression analysis considering a set of household characteristics.\footnote{The mean value of the MPC in our sample equals to 43.5 percent, which is close to the average percentages reported in \citet{Drescher2020} and \citet{Albacete2025} using the third wave of the HFCS. Across countries, MPCs range from 29.2 in Portugal to 55 in Croatia and Lithuania. The estimate for Italy is also close to those reported in previous studies \citep{Jappelli2014, Jappelli2020}. These average reported MPCs are larger than current estimates based on survey data on observed consumption and income changes. As is frequently the case with subjective perceptions and expectations data \citep{Manski2018, Bruine2022, Kosar2023, Arellano2024}, our variable concerning the MPC is clustered around specific rounded values like 0 percent (saving all), 50 percent (saving/spending half), and 100 percent (spending all), similar to \citet{Albacete2025}. In addition, the strong buching responses at 0/50/100 tend to be some persistent one and two periods later. These features are displayed in Tables~\ref{Table::MPC}-\ref{Table::MPC_transition} and Figures~\ref{Fig::MPC_distribution}-\ref{Fig::MPC_change_distribution} in Appendix~\ref{Appendix::B}.} Specifically, we estimate the following baseline linear regression model for the MPC by pooled Ordinary Least Squares (OLS):
\begin{align}\label{Eq::estimable}
\begin{split}
MPC_{ijt}=\alpha_0+\beta_1 Poor~HTM_{ijt}+\beta_2 Wealthy~HTM_{ijt}+ \boldsymbol{X_{ijt}^\prime} \boldsymbol{\gamma}+\mu_j+\tau_t+\varepsilon_{ijt},
\end{split}
\end{align} 
where subscript $i$ refers to the household $i$, subscript $j$ denotes the country of residence $j$, and subscript $t$ represents the survey year $t$. The dependent variable, $MPC_{ijt}$, is the MPC reported by household $i$ residing in country $j$ in period $t$, whereas $Poor~HTM_{ijt}$ and $Wealthy~HTM_{ijt}$, our two key explanatory variables in Equation~\eqref{Eq::estimable}, are two dummy variables each taking value 1 for households who are either poor or wealthy HtM at time $t$, respectively, and value 0 otherwise. The $\beta_k$ coefficients, for $k=1,2$, are the main parameters of interest from Eq.~\eqref{Eq::estimable} and measure the percentage point differences in the MPC across poor and wealthy HtM households, relative to their non-HtM household counterparts.

On the other hand, $X_{ijt}$ is a vector of household-level control characteristics, which includes households' heterogeneous demographic characteristics such as the head of household's age, indicators for household heads who are males, their education levels and employment status, the family structure (e.g., marital status, household size and number of children), and the specific housing tenure status through two dummy variables denoting mortgagors and renters, with outright owners and others considered as the baseline category for housing tenure status.\footnote{We control for age using 5-year intervals to capture potential non-linearities in consumption \citep{Kaplan2014}. Our results are robust to the inclusion of log family income and a set of country-year fixed effects.} Finally, the terms $\tau_t$ and $\mu_j$ indicate wave and country fixed effects, and $\varepsilon_{ijt}$ is an error term.\footnote{An alternative estimator would be a Tobit model \citep{Jappelli2014, Bunn2018, Christelis2021, Christelis2025, Albuquerque2023, Albacete2025}, as MPCs are constrained to vary from 0 to 100. However, this estimator is not compatible with bootstrapped replicates used in the HFCS.}

%%%%%%%%%%%%%%%%%%%%%%%%%%%%%%%%%%%%%%%%
%%%	Results
%%%%%%%%%%%%%%%%%%%%%%%%%%%%%%%%%%%%%%%%

\section{Results}\label{Sec::MPC}

Table~\ref{tab::OLS_Results} presents the results of estimating Equation~\eqref{Eq::estimable}.  In Column (1) we find that HtM households do not display higher MPCs, according to the definition in KVW. In contrast, HtM households in net worth \citep{Zeldes1989} show higher MPCs in Column (2), relative to their non-HtM counterparts. Quantitatively, a HtM household in net worth displays 2.441 more percentage points in the MPC than non-HtM households. 

Turning to our main estimates from Eq.~\eqref{Eq::estimable} shown in Column (3), we find a positive association between the MPC and poor HtM status, and a negative relationship with wealthy HtM households. In particular, we find that poor HtM households display the highest MPC, and being a poor HtM household correlates to an increase by about 6.029 percentage points in the MPC, relative to non-HtM households. That is, these types of households, who ultimately hold litle wealth, consume the majority of the extra liquidity they receive and display the highest MPC, in line with theoretical predictions of heterogeneous MPCs when people face binding liquidity constraints that prevent them from smoothing consumption fluctuations in response to income changes. On the other hand, wealthy HtM households have MPCs that are 1.789 percentage points lower than those of non-HtM households, a magnitude which is statistically significant at the 1\% level. This result implies that the shock is not large enough to relax the financial constraint, which may be a reasonable assumption given the survey question we exploit, and that these households would use part of the income shock to increase their liquid savings due to a precautionary motive, pay off their debts or adjust their balance sheets, rather than spending. Unfortunately, the survey question does not allow us to distinguish among these channels and households who repay debt, similar to \citet{Jappelli2014, Jappelli2020} and \citet{Jappelli2026}. 

As a result, despite their limited holdings of liquid assets, wealthy HtM households display lower MPCs than the non-HtM. This underscores the composition of household resources for spending responses, as spending is more responsive to liquid assets rather than total wealth \citep{Kaplan2014}, and adds a new result in the related literature that finds a negative correlation between MPC and liquid wealth or cash-on-hand \citep{Jappelli2014, Jappelli2020, Bunn2018, Christelis2019, Christelis2021, Fuster2021, Albacete2025}.\footnote{Using the 2017 HFCS, \citet{Drescher2020} did not find any relationship between net wealth and MPCs. However, they did not analyze the MPC across poor, wealthy, and non-HtM households.} A potential explanation for the above result arises from the significant holdings of illiquid assets across the wealthy HtM that cannot be used immediately to smooth consumption in the event of shocks, meaning that they may differ in their spending purposes from those who do not hold neither liquid nor illiquid assets (i.e., poor HtM households). 

Overall, both subsets of households fail to smooth consumption in response to transitory increases in income. This result is contrary to the theory that suggests that the MPC out of a transitory and small income shock will either be relatively small or very close to zero (i.e., a transitory change in disposable income should not trigger meaningful responses in current consumption, whereas the response should be smaller for small shocks in comparison to large ones), and aligns with life-cycle models that incorporate liquidity constraints and precautionary saving purposes to explain the lack of consumption smoothing of agents and high MPCs. 

Turning to the rest of our estimates in Table~\ref{tab::OLS_Results}, we find that other observable household characteristics are associated with MPC heterogeneity. Specifically, we find that older individuals tend to have higher MPCs relative to younger groups, meaning that the MPC increases with age, particularly for the oldest age group. That is, the income windfall is not transmitted to savings among the elderly. Specifically, those in the oldest age group (75 and over) exhibit the largest MPCs compared to the youngest group (less than 25 years old, the reference category). These older households have relatively fewer periods than young over to spread changes in lifetime income and smooth their consumption so this is a result consistent with the predictions of standard life-cycle models with finite planning horizons and no bequest motives for saving \citep{ModiglianiBrumberg1954}.\footnote{These older households may leave inheritances at the end, but these are involuntary and unplanned.} 

We also find that the education level attainment correlates with the MPC, and secondary as well as university educated households appear to display a significantly larger MPC than those with primary or no education, in contrast to \citet{Blundell2008}. Specifically, secondary educated household heads have MPCs 1.98-2.14 percentage points higher than primary educated household heads, whereas university educated households heads display 1.86 more percentage points in the MPC than primary educated household heads.\footnote{We alternatively distinguish households with and without college education, similar to \citet{Blundell2008}, and do not find that tertiary educated household heads display lower MPCs, in comparison to less educated households.} On the other hand, we find lower MPCs for employed household heads and mortgagors as well as renters. The empirical findings for employed, mortgagor, and renter households are consistent with precautionary saving models and frameworks incorporating liquidity constraints. Unemployed individuals, who face greater income volatility, tend to accumulate larger precationary buffers relative to their employed counterparts. On the other hand, mortgagor households face tighter liquidity constraints arising from debt service obligations, whereas renters are on average younger households. As a result, the transitory income windfall is more likely to be saved or applied to debt service rather than consumed among these. Quantitatively, households headed by an employed individual exhibit MPCs approximately 4.4--4.7 percentage points lower than those of non-working households, whereas mortgagors and renters report MPCs that are lower by approximately 3.06 and 1.81 percentage points, respectively, relative to outright owners and households in other primary tenure categories. Most of these relationships have been reported in previous works \citep{Jappelli2014, Jappelli2020, Bunn2018, Christelis2019, Christelis2021, Albuquerque2023, Sokolova2023, Albacete2025}.

\begin{table}[H]
\begin{center}
\caption{Pooled OLS regression results}\label{tab::OLS_Results}
\resizebox{0.5\textwidth}{!}{
\begin{tabular}{lccc}
\toprule
 & (1) & (2) & (3) \\ \midrule
HtM & -0.050 & --- & --- \\
 & (0.595) & & \\
HtM in net worth & --- & 2.441$^{**}$ & --- \\
 & & (1.123) & \\
Poor HtM & --- & --- & 6.029$^{***}$ \\
 & & & (1.412) \\
Wealthy HtM & --- & --- & -1.789$^{***}$ \\
 & & & (0.574) \\
Male & -0.042 & -0.050 & -0.105 \\
 & (0.469) & (0.469) & (0.466) \\
25 -- 29 years & 0.468 & 0.407 & 0.295 \\
 & (3.279) & (3.228) & (3.212) \\
30 -- 34 years & 1.692 & 1.640 & 1.472 \\
 & (3.163) & (3.119) & (3.095) \\
35 -- 39 years & 3.553 & 3.566 & 3.471 \\
 & (3.106) & (3.077) & (3.057) \\
40 -- 44 years & 4.490 & 4.472 & 4.253 \\
 & (3.051) & (3.017) & (2.998) \\
45 -- 49 years & 5.761$^{*}$ & 5.752$^{*}$ & 5.575$^{*}$ \\
 & (3.069) & (3.015) & (2.976) \\
50 -- 54 years & 5.539$^{*}$ & 5.558$^{*}$ & 5.315$^{*}$ \\
 & (3.078) & (3.044) & (3.014) \\
55 -- 59 years & 6.162$^{**}$ & 6.190$^{**}$ & 5.939$^{**}$ \\
 & (2.993) & (2.956) & (2.932) \\
60 -- 64 years & 5.292$^{*}$ & 5.384$^{*}$ & 5.196$^{*}$ \\
 & (2.990) & (2.961) & (2.932) \\
65 -- 69 years & 5.845$^{*}$ & 5.994$^{**}$ & 5.817$^{**}$ \\
 & (3.009) & (2.971) & (2.934) \\
70 -- 74 years & 5.049$^{*}$ & 5.249$^{*}$ & 5.100$^{*}$ \\
 & (3.054) & (3.007) & (2.978) \\
75 -- 79 years & 7.709$^{**}$ & 7.932$^{***}$ & 7.765$^{***}$ \\
 & (3.013) & (2.967) & (2.940) \\
Secondary education & 1.978$^{*}$ & 2.099$^{**}$ & 2.137$^{**}$ \\
 & (1.072) & (1.026) & (0.978) \\
Tertiary education & 1.645 & 1.881$^{*}$ & 1.860$^{**}$ \\
 & (1.016) & (0.979) & (0.944) \\
Employed & -4.736$^{***}$ & -4.611$^{***}$ & -4.392$^{***}$ \\
 & (0.717) & (0.684) & (0.653) \\
Married & 0.789 & 0.825 & 0.837 \\
 & (0.529) & (0.525) & (0.524) \\
Household size & 0.104 & 0.106 & 0.125 \\
 & (0.298) & (0.294) & (0.284) \\
Number of children & 0.120 & 0.074 & 0.050 \\
 & (0.426) & (0.425) & (0.417) \\
Mortgagor & -3.041$^{***}$ & -3.067$^{***}$ & -3.055$^{***}$ \\
 & (0.598) & (0.587) & (0.584) \\
Renter & -1.029$^{*}$ & -1.456$^{**}$ & -1.809$^{***}$ \\
 & (0.612) & (0.659) & (0.657) \\
Constant & 31.148$^{***}$ & 30.869$^{***}$ & 31.352$^{***}$ \\
& (3.230) & (3.188) & (3.167) \\
 & & & \\
Country fixed effects & Yes & Yes & Yes \\
Year fixed effects & Yes & Yes & Yes \\
Observations & 157,430 & 157,430 & 157,430 \\
\bottomrule
\end{tabular} }
\begin{minipage}{0.5\textwidth}
\footnotesize
\textit{Notes}: OLS estimates. Standard errors in parentheses. Data come from the Household Finance and Consumption Survey, 2017--2023. Sample is restricted to households whose head is between 22 and 79 years old, have no negative family income and whose income is not entirely derived from self-employment. Estimates are weighted using sampling and bootstrapped replicated weights. $^{*}$ $p<$ 0.10, $^{**}$ $p<$ 0.05, $^{***}$ $p<$ 0.01.
\end{minipage}
\end{center}
\end{table}

Given that the HFCS has a panel component over the period 2017--2023, we also employ a panel data regression model and rerun Equation~\eqref{Eq::estimable} further incorporating household fixed effects (e.g. a $\alpha_i$ term in Eq.~\eqref{Eq::estimable}) that initially remain in the error term and account for household unobserved heterogeneity that is fixed over time, such as personality traits, risk aversion or discount rates. These features might affect both HtM behavior and consumption decisions simultaneously \citep{Jappelli2020, Aguiar2025}, thus introducing a potential bias in the OLS estimates of $\beta_1$ and $\beta_2$ in Eq.~\eqref{Eq::estimable}, which exploits variation both between and within households. (Hausman test results support the validity of the fixed effects specification).

This approach is similar to that taken by \citet{Jappelli2020}, who also have panel data on a sample of Italian households over two years in 2010 and 2016.\footnote{As argued in \citet{Jappelli2020}, in contrast to \citet{Bunn2018}, \citet{Christelis2019}, \citet{Fuster2021} and \citet{Jappelli2026}, we have data for the same households at (at least) two different points in time, rather than on a different set of MPC questions asked at the same point. For further details on this matter, we refer to footnote 9 in their paper.}

To properly include household fixed effects, we need households to be observed at least twice to identify coefficients. To do so, we use the panel information of the HFCS over 2017--2023 and focus on a sample of household heads observed for at least two periods, which restricts our analysis to an unbalanced panel sample of 63,602 household-year observations from 26,814 households.\footnote{In principle, the analysis requires at least two observations per household, either consecutive or non-consecutive. However, we focus on households observed for at least two \textit{consecutive} waves due to the HFCS design of the longitudinal household identifier.}

Table~\ref{tab::FE_Results} shows the results of estimating Equation~\eqref{Eq::estimable} after incorporating household fixed effects.\footnote{In principle, household time-invariant variables such as gender or educational level attainment should be excluded. However, the identity of the household head could change between interviews in the HFCS. Household fixed effects subsume country fixed effects.} These estimates are comparable to those in \citet{Jappelli2020}. We find that controlling for household unobserved heterogeneity significantly changes the main results, and previous pooled OLS estimates exagerate the negative relationship between wealthy HtM and MPC as well as the positive relationship with poor HtM. Specifically, the coefficient estimated for poor HtM households is no longer statistically significant after controlling for unobserved heterogeneity, whereas the negative estimate for wealthy HtM household remains statistically significant (at the 1\% level), albeit slightly smaller in absolute terms.\footnote{Poor HtM households are relatively persistent over time, which may reflect the insignificant estimated effect within this group. Nevertheless, sufficient within-household variation exists to identify the coefficient.} The comparison between pooled OLS and household fixed effects results for poor and wealthy HtM households fits the patterns displayed by \citet{Jappelli2020}, who also find that OLS estimates exaggerates the negative relationship between the cash-on-hand distribution and MPC \citep{Jappelli2020}, meaning that unobserved factors also account for the observed relationship in the cross-section. The remaining statistically significant variables are the dummy variables for employed household heads, mortgagors and renters, which indicates that households headed by an employed individual exhibit MPCs approximately 2.5 percentage points lower than those headed by a non-employed individual, while mortgagor and renter households report MPCs approximately 1.8 and 3.2 percentage points lower than those of outright owners and households with other housing tenure statuses, respectively.

\begin{table}[H]
\begin{center}
\caption{Fixed effects regression results}\label{tab::FE_Results}
\resizebox{0.5\textwidth}{!}{
\begin{tabular}{lccc}
\toprule
 & (1) & (2) & (3) \\ \midrule
HtM & -1.326$^{***}$ & --- & --- \\
 & (0.502) & & \\
HtM in net worth & --- & 0.655 & --- \\
 & & (1.137) & \\
Poor HtM & --- & --- & 1.049 \\
 & & & (1.361) \\
Wealthy HtM & --- & --- & -1.671$^{***}$ \\
 & & & (0.524) \\
Male & 1.544$^{*}$ & 1.521$^{*}$ & 1.544$^{*}$ \\
 & (0.819) & (0.819) & (0.819) \\
25 -- 29 years & 1.168 & 1.192 & 1.206 \\
 & (2.803) & (2.812) & (2.801) \\
30 -- 34 years & 3.699 & 3.746 & 3.736 \\
 & (2.954) & (2.963) & (2.951) \\
35 -- 39 years & 2.063 & 2.125 & 2.130 \\
 & (3.035) & (3.043) & (3.032) \\
40 -- 44 years & 1.101 & 1.189 & 1.173 \\
 & (3.089) & (3.096) & (3.086) \\
45 -- 49 years & 1.191 & 1.238 & 1.242 \\
 & (3.120) & (3.128) & (3.117) \\
50 -- 54 years & 0.103 & 0.136 & 0.130 \\
 & (3.152) & (3.160) & (3.148) \\
55 -- 59 years & 0.522 & 0.536 & 0.556 \\
 & (3.206) & (3.212) & (3.201) \\
60 -- 64 years & 0.188 & 0.214 & 0.231 \\
 & (3.281) & (3.288) & (3.276) \\
65 -- 69 years & 1.185 & 1.234 & 1.235 \\
 & (3.371) & (3.378) & (3.366) \\
70 -- 74 years & 0.862 & 0.914 & 0.915 \\
 & (3.491) & (3.500) & (3.486) \\
75 -- 79 years & 1.610 & 1.657 & 1.654 \\
 & (3.642) & (3.650) & (3.637) \\
Secondary education & -0.531 & -0.512 & -0.530 \\
 & (1.313) & (1.313) & (1.311) \\
Tertiary education & -1.111 & -1.092 & -1.113 \\
 & (1.607) & (1.607) & (1.606) \\
Employed & -2.525$^{***}$ & -2.506$^{***}$ & -2.497$^{***}$ \\
 & (0.668) & (0.667) & (0.667) \\
Married & -0.434 & -0.433 & -0.420 \\
 & (0.937) & (0.937) & (0.937) \\
Household size & -0.536 & -0.519 & -0.536 \\
 & (0.413) & (0.413) & (0.413) \\
Number of children & 0.985$^{*}$ & 0.969$^{*}$ & 0.981$^{*}$ \\
 & (0.569) & (0.569) & (0.569) \\
Mortgagor & -1.846$^{**}$ & -1.805$^{**}$ & -1.858$^{**}$ \\
 & (0.788) & (0.788) & (0.788) \\
Renter & -3.098$^{***}$ & -3.107$^{***}$ & -3.214$^{***}$ \\
 & (1.179) & (1.180) & (1.180) \\
Constant & 45.955$^{***}$ & 45.591$^{***}$ & 45.893$^{***}$ \\
& (3.532) & (3.540) & (3.528) \\
 & & & \\
Year fixed effects & Yes & Yes & Yes \\
Observations & 63,602 & 63,602 & 63,602 \\
Households & 26,814 & 26,814 & 26,814 \\
\bottomrule
\end{tabular} }
\begin{minipage}{0.5\textwidth}
\footnotesize
\textit{Notes}: FE estimates. Robust standard errors clustered at the household level in parentheses. Data come from the Household Finance and Consumption Survey, 2017--2023. Sample is restricted to households whose head is between 22 and 79 years old, have no negative family income and whose income is not entirely derived from self-employment, observed for at least two periods. $^{*}$ $p<$ 0.10, $^{**}$ $p<$ 0.05, $^{***}$ $p<$ 0.01.
\end{minipage}
\end{center}
\end{table}

In Table~\ref{tab::Robustness} we show that our main results are robust to different robustness checks involving alternative subsamples or model specifications. Mimicking \citet{Jappelli2014, Jappelli2020}, in Column (1), we rerun our Eq.~\eqref{Eq::estimable} after excluding those who declare MPCs equal to the 50 value, which may indicate respondent indecisiveness and measurement error, whereas in Column (2) we only keep those households who declare that their spending during the year was similar to what they would do in a ``normal'' year \citep{Castaldo2025}.\footnote{The question is: ``\textit{Aside from any purchases of assets, would you say that your household's regular expenses over the last 12 months were higher than normal, lower than normal, or were they about normal?}''. We keep households who report that their consumption in 2021 was ``\textit{just about normal}''.} The reported MPC could depend on the macroeconomic context \citep{Jappelli2014}. In our unbalanced panel, the sample period includes the year immediately following the Covid-19 pandemic, which may bias the MPC estimates, influence HtM transitions and limit the external validity of our findings, even though the most common reference period was not 2020 and ultimately conducted throughout 2021. In Column (3) we include an indicator for those who declare a loan rejection or were discouraged from applying for credit in the past 12 months, an important control according to previous estimates on MPCs \citep{Bunn2018, Jappelli2020}, along with total non-durable consumption (in logarithmic terms) as our proxy for permanent income \citep{Jappelli2020, Boehm2025}. Fourth, in Column (4), we restrict the sample to household heads aged 25--64 \citep{Aguiar2025} to minimize the influence of major life-cycle decisions, such as retirement. Besides, many of the variables considered, such as housing tenure, seem closely related to the HtM status of a household. However, removing this control variable does not affect the results in Column (5) of Table~\ref{tab::Robustness}. Another issue is that the estimates for individual variables, such as the negative statistically significant coefficient estimated for employed households, could reflect changes in the household head's identity. Nevertheless, restricting to same household heads with no changes in marital status \citep{Blundell2008, Jappelli2020} does not alter our main findings in Column (6).\footnote{We emphasize that this does not impact household variables, such as the dummy variables denoting HtM household status.} For the sake of simplicity, we only report the significant coefficients estimated in Table~\ref{tab::Robustness}, which are robust and very similar.\footnote{We explore other alternatives for the structure of the error term, like country level clustered standard errors, and find identical conclusions.}

\begin{landscape}
\begin{table}[H]
\begin{center}
\caption{Robustness checks, fixed effects regression results}\label{tab::Robustness}
\resizebox{1.5\textwidth}{!}{
\begin{tabular}{lcccccc}
\toprule
 & No MPCs equal to 50 & ``Normal'' consumption in 2021 & Additional controls & Heads aged 25--64 & No housing tenure variables & Same household head \\ \midrule
Poor HtM & 2.206 & 0.222 & 0.659 & 0.881 & 0.961 & 0.872 \\
 & (1.819) & (1.550) & (1.381) & (1.541) & (1.362) & (1.597) \\
Wealthy HtM & -1.733$^{**}$ & -2.059$^{***}$ & -1.676$^{***}$ & -1.393$^{**}$ & -1.631$^{***}$ & -1.394$^{**}$ \\
 & (0.750) & (0.630) & (0.529) & (0.614) & (0.523) & (0.594) \\
Employed & -2.858$^{***}$ & -2.432$^{***}$ & -2.539$^{***}$ & -2.661$^{***}$ & -2.545$^{***}$ & -2.030$^{**}$ \\
& (0.967) & (0.768) & (0.674) & (0.806) & (0.667) & (0.797) \\
Mortgagor & -2.202$^{*}$ & -1.287 & -1.792$^{**}$ & -1.285 & --- & -1.730$^{**}$ \\
 & (1.143) & (0.921) & (0.799) & (0.906) & & (0.870) \\
Renter & -3.027$^{*}$ & -3.043$^{**}$ & -3.022$^{**}$ & -3.051$^{**}$ & --- & -3.501$^{***}$ \\
 & (1.679) & (1.344) & (1.188) & (1.355) & & (1.319) \\
Credit constraints &  --- & --- & 0.087 & --- & --- & --- \\
& & & (0.885) & & & \\
Permanent income & --- & --- & 0.794$^{**}$ & --- & --- & --- \\
 & & & (0.332) & & & \\
 & & & & & & \\
Rest of controls from Eq.~\eqref{Eq::estimable}& Yes & Yes & Yes & Yes & Yes & Yes \\
Year fixed effects & Yes & Yes & Yes & Yes & Yes & Yes \\
Observations & 39,261 & 46,378 & 62,844 & 40,648 & 63,602 & 52,088 \\
Households & 17,435 & 20,312 & 26,591 & 17,461 & 26,814 & 22,355 \\
\bottomrule
\end{tabular} }
\begin{minipage}{1.5\textwidth}
\footnotesize
\textit{Notes}: FE estimates. Robust standard errors clustered at the household level in parentheses. Data come from the Household Finance and Consumption Survey, 2017--2023. Sample is restricted to households whose head is between 22 and 79 years old, have no negative family income and whose income is not entirely derived from self-employment, observed for at least two periods. $^{*}$ $p<$ 0.10, $^{**}$ $p<$ 0.05, $^{***}$ $p<$ 0.01.
\end{minipage}
\end{center}
\end{table}
\end{landscape}

Given the harmonized cross-country design of the survey, in Table~\ref{tab::CountryGroups} we analyze the relationship between HtM and MPC behavior across country groups. Specifically, we group countries into Western (Belgium, Germany, France, Ireland, Netherlands), Southern (Cyprus, Spain, Italy, Malta), and Eastern (Estonia, Lithuania, Latvia, Slovakia) European countries.\footnote{Limited sample sizes within countries prevent a detailed analysis at the country level. Table~\ref{tab::Composition} reports the country-level sample composition of the unbalanced panel.} We find consistent results across all these country groups.

Finally, we examine coarse responses and focus on the fraction of households reporting MPCs of 0, 50, and 100 percent in Table~\ref{tab::LPM_Results}. To do so, we estimate linear probability models with household fixed effects for three separate binary outcomes, each taking the value 1 when the reported MPC equals 0, 50, and 100 percent, respectively. We find that wealthy HtM households are 2.4 percentage points more likely to save the additional income. In contrast, poor HtM households are 5.7 percentage points more likely than non-HtM households to spend the entire income windfall, consistent with life-cycle models incorporating binding financial constraints. Besides, they are also less likely to report an MPC of 50 percent. This latter finding may reflect either behavioral differences or a lower degree of uncertainty in the interpretation of the survey question \citep{Jappelli2020}.

\begin{table}[H]
\begin{center}
\caption{Fixed effects regression results across country groups}\label{tab::CountryGroups}
\resizebox{0.5\textwidth}{!}{
\begin{tabular}{lccc}
\toprule
 & Western & Southern & Eastern \\ \midrule
Poor HtM & 0.819 & 0.695 & 4.265 \\
 & (2.131) & (1.862) & (5.180) \\
Wealthy HtM & -1.153$^{*}$ & -1.830$^{**}$ & -4.296$^{**}$ \\
 & (0.680) & (0.930) & (1.768) \\
Male & 0.547 & 2.275$^{*}$ & 2.457 \\
 & (1.236) & (1.223) & (2.500) \\
25 -- 29 years & -0.110 & 0.885 & 9.528 \\
 & (3.209) & (6.604) & (9.525) \\
30 -- 34 years & 3.571 & 6.295 & 1.769 \\
 & (3.482) & (6.545) & (9.813) \\
35 -- 39 years & 2.562 & 3.376 & -0.245 \\
 & (3.682) & (6.532) & (9.921) \\
40 -- 44 years & 1.826 & 1.550 & 0.365 \\
 & (3.799) & (6.519) & (10.013) \\
45 -- 49 years & 1.776 & 1.120 & 3.353 \\
 & (3.911) & (6.462) & (9.947) \\
50 -- 54 years & -0.269 & 1.137 & 2.444 \\
 & (3.991) & (6.467) & (9.980) \\
55 -- 59 years & -0.015 & 1.344 & 4.444 \\
 & (4.122) & (6.471) & (10.126) \\
60 -- 64 years & 0.245 & 0.216 & 3.103 \\
 & (4.275) & (6.526) & (10.390) \\
65 -- 69 years & 1.102 & 1.264 & 4.407 \\
 & (4.446) & (6.616) & (10.642) \\
70 -- 74 years & -0.380 & 2.439 & 4.088 \\
 & (4.685) & (6.719) & (10.992) \\
75 -- 79 years & 0.829 & 2.977 & 1.626 \\
 & (4.989) & (6.876) & (11.488) \\
Secondary education & 3.546 & -1.418 & -25.207$^{***}$ \\
 & (2.764) & (1.505) & (9.149) \\
Tertiary education & 1.514 & -0.373 & -26.006$^{***}$ \\
 & (3.014) & (2.128) & (9.129) \\
Employed & -3.603$^{***}$ & -2.140$^{**}$ & 0.510 \\
 & (1.001) & (0.996) & (2.037) \\
Married & -0.010 & -0.133 & -3.735 \\
 & (1.308) & (1.537) & (2.907) \\
Household size & -1.013 & 0.106 & -1.335 \\
 & (0.641) & (0.605) & (1.308) \\
Number of children & 1.570$^{**}$ & -0.112 & 2.705 \\
 & (0.764) & (0.954) & (1.907) \\
Mortgagor & -1.566 & -2.248$^{*}$ & -3.875 \\
 & (1.068) & (1.283) & (3.310) \\
Renter & -4.641$^{***}$ & 0.037 & -6.795 \\
 & (1.580) & (1.942) & (5.203) \\
Constant & 44.724$^{***}$ & 42.965$^{***}$ & 73.204$^{***}$ \\
& (5.009) & (6.755) & (13.938) \\
 & & & \\
Year fixed effects & Yes & Yes & Yes \\
Observations & 32,374 & 24,868 & 6,360 \\
Households & 13,301 & 10,352 & 3,161 \\
\bottomrule
\end{tabular} }
\begin{minipage}{0.5\textwidth}
\footnotesize
\textit{Notes}: FE estimates. Robust standard errors clustered at the household level in parentheses. Data come from the Household Finance and Consumption Survey, 2017--2023. Sample is restricted to households whose head is between 22 and 79 years old, have no negative family income and whose income is not entirely derived from self-employment, observed for at least two periods. $^{*}$ $p<$ 0.10, $^{**}$ $p<$ 0.05, $^{***}$ $p<$ 0.01.
\end{minipage}
\end{center}
\end{table}

\begin{table}[H]
\begin{center}
\caption{Linear probability model results}\label{tab::LPM_Results}
\resizebox{0.9\textwidth}{!}{
\begin{tabular}{lccc}
\toprule
 & MPC equal to 0 & MPC equal to 50 & MPC equal to 100 \\ \midrule
Poor HtM & 0.022 & -0.038$^{**}$ & 0.057$^{***}$ \\
 & (0.016) & (0.015) & (0.016) \\
Wealthy HtM & 0.024$^{***}$ & -0.004 & -0.007 \\
 & (0.007) & (0.007) & (0.006) \\
Employed & 0.030$^{***}$ & -0.007 & -0.014$^{*}$ \\
 & (0.008) & (0.009) & (0.007) \\
Mortgagor & 0.032$^{***}$ & -0.005 & -0.006 \\
 & (0.010) & (0.010) & (0.009) \\
 & & \\
Rest of controls from Eq.~\eqref{Eq::estimable}& Yes & Yes & Yes \\
Year fixed effects & Yes & Yes & Yes \\
Observations & 63,602 & 63,602 & 63,602 \\
Households & 26,814 & 26,814 & 26,814 \\
\bottomrule
\end{tabular} }
\begin{minipage}{0.9\textwidth}
\footnotesize
\textit{Notes}: Linear probability model with FE estimates. Robust standard errors clustered at the household level in parentheses. Data come from the Household Finance and Consumption Survey, 2017--2023. Sample is restricted to households whose head is between 22 and 79 years old, have no negative family income and whose income is not entirely derived from self-employment, observed for at least two periods. $^{*}$ $p<$ 0.10, $^{**}$ $p<$ 0.05, $^{***}$ $p<$ 0.01.
\end{minipage}
\end{center}
\end{table}

%%%%%%%%%%%%%%%%%%%%%%%%%%%%%%%%%%%%%%%%
%%%	Conclusions
%%%%%%%%%%%%%%%%%%%%%%%%%%%%%%%%%%%%%%%%

\section{Conclusions}\label{Sec::Conclusions}

A key parameter driving the excess sensitivity of consumption to income changes is liquidity constraints. This paper studies the consumption responses to income changes of HtM households in Europe. To do so, it uses harmonized cross-country data from a total of 23 countries of the HFCS over 2010--2023 and examines the MPC across these groups through a hypothetical survey question that captures how households believe they would modify their consumption in response to a transitory, unexpected, income gain. Related literature has typically used semistructural methods based on statistical decomposition and covariance restrictions on the joint distribution of income and consumption growth, which requires very long panels along with information on all the components of the budget constraint rarely available.

We find wide cross-country heterogeneity and that the shares of HtM households oscillate from 13.9\% in Czech Republic to 70.6\% in Greece. In addition, the vast majority of HtM households are wealthy in Europe, over 60\% in all countries analyzed, indicating that most of those households hold positive, and sometimes substantial, amounts of illiquid wealth. This result is consistent with previous works in either developed or developing economies. Specifically, the percentage of HtM that are considered wealthy HtM ranges from 64\% in Ireland and Italy to 90\% in Greece, Luxembourg and Malta. Concerning MPCs, poor HtM households exhibit the highest MPC, as expected due to their limited access to both liquid and illiquid assets, whereas wealthy HtM households have lower MPCs compared to their non-HtM counterparts. 

The dataset further enables us to analyze potential biases arising from cross-sectional regressions on the MPC, as it follows a subset of households over time which allows to focus on households who switch their HtM status over time. Interestingly, we show that pooled OLS estimates exaggerate the estimates for poor and wealthy HtM households. Specifically, the initial positive association reported for poor HtM households is no longer statistically significant once household unobserved heterogeneity is accounted for. Similarly, the estimated relationship for wealthy HtM households remains negative and statistically significant, although it is lower in absolute terms after controlling for cofounders and household unobserved heterogeneity. Finally, we show that poor HtM households are more likely to spend the whole windfall gain. All these results align with life-cycle models with liquidity constraints and precationary saving motives.

Overall, these MPC results highlight the joint analysis of liquid and illiquid assets in the household balance sheets to identify constrained households and study the excess sensitivity in consumption to temporary changes in income. As expected, the fraction of poor HtM households correlates positively with the MPC. However, the heterogeneity shown within HtM households suggests that liquid wealth may be an imperfect proxy to define financial constrained status across households.

This paper offers new results on the fraction of HtM households and their consumption behavior using harmonized cross-country data in Europe. Despite the important findings, we are aware of several limitations arising from this work that may motivate future works. On the one hand, the variable concerning the MPC only considers hypothetical windfall gains and consumption/saving choices on the short-run. The sign of income shocks is an important determinant of the MPC, whereas the windfall gain may induce debt repayments or portfolio allocation, in addition to spending or saving. On the other, similar to other methodological alternatives, limitations concerning this reported approach to estimating MPCs include its external validity and quality of data, despite recent work pointing to the validity of this approach.

\clearpage

\begin{singlespace}

\end{singlespace}

%%%%%%%%%%%%%%%%%%%%%%%%%%%%%%%%%%%%%%%%
%%% Appendix settings
%%%%%%%%%%%%%%%%%%%%%%%%%%%%%%%%%%%%%%%%

% The appendix command is issued once, prior to all appendices, if any.
\let\origappendix\appendix % save the existing appendix command
\renewcommand\appendix{\clearpage\pagenumbering{arabic}\origappendix}
\appendix
\renewcommand\appendixtocname{Appendix}
\renewcommand\appendixpagename{Appendix}
\appendixpage           % Adds an 'Appendix' title before the first appendix
\addappheadtotoc        % Adds an 'Appendix' title in the table of contents
%   Separate enumeration in the appendix:
\numberwithin{equation}{section} 
\numberwithin{table}{section}   
\numberwithin{figure}{section}
\setcounter{footnote}{0}

%%%%%%%%%%%%%%%%%%%%%%%%%%%%%%%%%%%%%%%%
%%% Appendix A: HtM
%%%%%%%%%%%%%%%%%%%%%%%%%%%%%%%%%%%%%%%%

\section{Additional results HtM}\label{Appendix::A}

\begin{table}[H]
\begin{center}
\caption{Comparison between HtM definitions} \label{Table::Additional_HtM}
\resizebox{0.8\textwidth}{!}{
\begin{tabular}{lccc}
\toprule
Country & Baseline & Zeldes in net worth & Zeldes in labor income \\
\midrule
Austria & 28.721 & 7.148 & 7.958 \\
Belgium & 26.045 & 6.589 & 6.297 \\
Cyprus & 43.280 & 11.218 & 11.339 \\
Czech Republic & 13.894 & 5.084 & 5.448 \\
Germany & 25.950 & 11.693 & 12.879 \\
Estonia & 22.990 & 6.528 & 6.382 \\
Spain & 23.326 & 10.133 & 12.062 \\
Finland & 30.100 & 11.636 & 12.893 \\
France & 27.864 & 8.731 & 8.791 \\
Greece & 70.634 & 10.275 & 11.471 \\
Hungary & 31.793 & 5.963 & 6.205 \\
Croatia & 41.187 & 6.594 & 7.576 \\
Ireland & 36.527 & 20.601 & 24.359 \\
Italy & 22.944 & 8.921 & 10.208 \\
Lithuania & 27.193 & 2.871 & 2.798 \\
Luxembourg & 26.012 & 6.452 & 7.491 \\
Latvia & 35.925 & 9.406 & 8.844 \\
Malta & 29.644 & 3.415 & 3.595 \\
Netherlands & 29.859 & 11.707 & 13.659 \\
Poland & 26.181 & 5.841 & 5.719 \\
Portugal & 25.967 & 8.937 & 9.400 \\
Slovenia & 37.484 & 8.306 & 8.764 \\
Slovakia & 26.329 & 3.698 & 3.604\\
\textbf{Pooled} & \textbf{27.286} & \textbf{9.515} & \textbf{10.434} \\
\bottomrule
\end{tabular} }
\begin{minipage}{0.8\textwidth}
\footnotesize
\textit{Notes}: Data come from the Household Finance and Consumption Survey, 2010--2023. Figures refer to sample averages, calculated using population weights and bootstrapped replicated weights. Column (3) focuses on 250,172 households with heads aged 25--64 years.
\end{minipage}
\end{center}
\end{table}
\clearpage

\begin{table}[H]
\begin{center}
\caption{HtM households, alternative definitions}\label{Table::Additional_HtM2}
\resizebox{1\textwidth}{!}{
\begin{tabular}{lccccccccccccccc}
\toprule
 & \multicolumn{3}{c}{Main criteria} & & \multicolumn{3}{c}{Pay period of 1 week} & & \multicolumn{3}{c}{Pay period of 1 month} & & \multicolumn{3}{c}{Other illiquid} \\
\cmidrule(lr){2-4} \cmidrule(lr){6-8} \cmidrule(lr){10-12} \cmidrule(lr){14-16}
Country & Wealthy & Poor & Total & & Wealthy & Poor & Total & & Wealthy & Poor & Total & & Wealthy & Poor & Total \\ \midrule
Austria & 24.200 & 4.521 & 28.721 & & 15.263 & 3.495 & 18.758 & & 39.289 & 5.970 & 45.260 & & 26.063 & 2.525 & 28.588 \\
Belgium & 21.833 & 4.212 & 26.045 & & 15.235 & 3.468 & 18.702 & & 33.347 & 5.123 & 38.471 & & 23.758 & 2.177 & 25.935 \\
Cyprus & 33.478 & 9.801 & 43.280 & & 30.731 & 9.167 & 39.897 & & 38.580 & 10.565 & 49.145 & & 40.461 & 2.742 & 43.203 \\
Czech Republic & 9.800 & 4.905 & 13.894 & & 7.658 & 3.551 & 11.209 & & 14.764 & 4.894 & 19.658 & & 11.669 & 2.157 & 13.827 \\
Germany & 18.584 & 7.366 & 25.950 & & 14.166 & 6.361 & 20.527 & & 28.263 & 8.740 & 37.003 & & 21.429 & 4.466 & 25.895 \\
Estonia & 18.173 & 4.818 & 22.991 & & 14.669 & 4.248 & 18.917 & & 24.306 & 5.612 & 29.918 & & 19.358 & 3.608 & 22.966 \\
Spain & 15.560 & 7.765 & 23.326 & & 13.369 & 7.059 & 20.427 & & 20.223 & 9.537 & 29.761 & & 19.582 & 3.738 & 23.321 \\
Finland & 22.521 & 7.579 & 30.100 & & 20.389 & 6.748 & 27.136 & & 26.847 & 9.146 & 35.993 & & 26.143 & 3.784 & 29.927 \\
France & 24.550 & 3.314 & 27.864 & & 17.736 & 2.352 & 20.088 & & 40.147 & 4.825 & 44.972 & & 27.782 & 0.128 & 27.911 \\
Greece & 65.325 & 5.309 & 70.634 & & 64.041 & 4.952 & 68.994 & & 67.843 & 5.901 & 73.744 & & 66.945 & 2.672 & 69.617 \\
Hungary & 26.929 & 4.864 & 31.793 & & 22.685 & 4.385 & 27.070 & & 34.365 & 5.816 & 40.181 & & 28.397 & 3.396 & 31.793 \\
Croatia & 35.377 & 5.809 & 41.187 & & 30.261 & 5.289 & 35.550 & & 43.724 & 6.882 & 50.606 & & 39.175 & 2.008 & 41.184 \\
Ireland & 23.671 & 12.856 & 36.527 & & 18.884 & 10.952 & 29.836 & & 30.709 & 15.387 & 46.096 & & 30.635 & 5.819 & 36.454 \\
Italy & 14.630 & 8.315 & 22.944 & & 13.402 & 7.714 & 21.115 & & 17.545 & 9.543 & 27.087 & & 21.453 & 1.521 & 22.974 \\
Lithuania & 24.874 & 2.319 & 27.193 & & 18.850 & 1.661 & 20.511 & & 36.621 & 3.091 & 39.712 & & 25.882 & 1.308 & 27.190 \\
Luxembourg & 22.050 & 3.962 & 26.012 & & 18.039 & 3.463 & 21.502 & & 29.754 & 5.100 & 34.853 & & 24.520 & 1.453 & 25.973 \\
Latvia & 28.740 & 7.185 & 35.925 & & 23.098 & 6.250 & 29.348 & & 36.926 & 8.981 & 45.906 & & 30.614 & 5.410 & 36.024 \\
Malta & 27.199 & 2.445 & 29.644 & & 25.229 & 2.135 & 27.364 & & 31.127 & 2.816 & 33.943 & & 29.077 & 0.760 & 29.837 \\
Netherlands & 25.089 & 4.770 & 29.859 & & 18.971 & 3.936 & 22.907 & & 36.251 & 6.183 & 42.431 & & 27.413 & 2.405 & 29.818 \\
Poland & 21.449 & 4.732 & 26.181 & & 17.499 & 4.236 & 21.735 & & 30.912 & 5.482 & 36.393 & & 22.803 & 3.378 & 26.181 \\
Portugal & 18.613 & 7.355 & 25.967 & & 14.195 & 6.132 & 20.328 & & 28.359 & 9.319 & 37.678 & & 22.105 & 3.769 & 25.874 \\
Slovenia & 30.281 & 7.202 & 37.484 & & 25.939 & 6.053 & 31.992 & & 37.844 & 8.433 & 46.277 & & 34.916 & 2.462 & 37.377 \\
Slovakia & 23.211 & 3.118 & 26.329 & & 18.333 & 2.511 & 20.844 & & 32.281 & 3.811 & 36.091 & & 24.525 & 1.755 & 26.280 \\
 & & & & & & & & & & & & & & & \\
\textbf{Pooled} & \textbf{21.115} & \textbf{6.172} & \textbf{27.286} & & \textbf{16.888} & \textbf{5.349} & \textbf{22.236} & & \textbf{30.087} & \textbf{7.529} & \textbf{37.617} & & \textbf{24.629} & \textbf{2.647} & \textbf{27.276} \\
\bottomrule
\end{tabular} }
\begin{minipage}{1\textwidth}
\footnotesize
\textit{Notes}: Data come from the Household Finance and Consumption Survey, 2010--2023. Figures refer to sample averages, calculated using population weights and bootstrapped replicated weights.
\end{minipage}
\end{center}
\end{table}
\clearpage

%%%%%%%%%%%%%%%%%%%%%%%%%%%%%%%%%%%%%%%%
%%% Appendix B: MPC
%%%%%%%%%%%%%%%%%%%%%%%%%%%%%%%%%%%%%%%%

\section{Additional results MPC}\label{Appendix::B}

\begin{table}[H]
\begin{center}
\caption{Average values for the MPC}\label{Table::MPC}
\resizebox{0.3\textwidth}{!}{
\begin{tabular}{lr}
\hline
Country & MPC \\
\hline
Austria & 47.220 \\
Belgium & 39.183 \\
Cyprus & 44.104 \\
Germany & 47.940 \\
Estonia & 37.797 \\
Spain & 39.821 \\
France & 38.773 \\
Greece & 51.676 \\
Croatia & 54.991 \\
Italy & 46.941 \\
Lithuania & 54.998 \\
Luxembourg & 34.422 \\
Latvia & 49.502 \\
Malta & 51.070 \\
Netherlands & 33.512 \\
Portugal & 29.205 \\
Slovenia & 48.065 \\
Slovakia & 54.298 \\
\textbf{Pooled} & \textbf{43.534} \\
\hline
\end{tabular} }
\caption*{\footnotesize \textit{Notes}: Data come from the Household Finance and Consumption Survey, 2017--2023.}
\end{center}
\end{table}
\clearpage

\begin{table}[H]
\begin{center}
\caption{Transition rates for  reported MPCs}\label{Table::MPC_transition}
\resizebox{\textwidth}{!}{
\begin{tabular}{lccc}
\toprule
 & MPC equal to 0 at ${t}$ & MPC equal to 50 at ${t}$ & MPC equal to 100 at ${t}$ \\ \midrule
\textit{Panel A: 21,345 households} & & \\
MPC equal to 0 at ${t+1}$ & 0.464 & 0.208 & 0.201 \\
MPC equal to 50 at ${t+1}$ & 0.202 & 0.376 & 0.247 \\ 
MPC equal to 100 at ${t+1}$ & 0.143 & 0.159 & 0.381 \\ \midrule
\textit{Panel B: 7,543 households} & & \\
MPC equal to 0 at ${t+2}$ & 0.435 & 0.212 & 0.223 \\
MPC equal to 50 at ${t+2}$ & 0.221 & 0.366 & 0.252 \\ 
MPC equal to 100 at ${t+2}$ & 0.155 & 0.168 & 0.340 \\
\bottomrule
\end{tabular} }
\begin{minipage}{\textwidth}
\footnotesize
\textit{Notes}: Data come from the Household Finance and Consumption Survey, 2017--2023.
\end{minipage}
\end{center}
\end{table}
\clearpage

\begin{figure}[H]
\centering

\caption{Histogram of the distribution on reported MPC}
\label{Fig::MPC_distribution}

\begin{threeparttable}
    \includegraphics[width=1\textwidth]{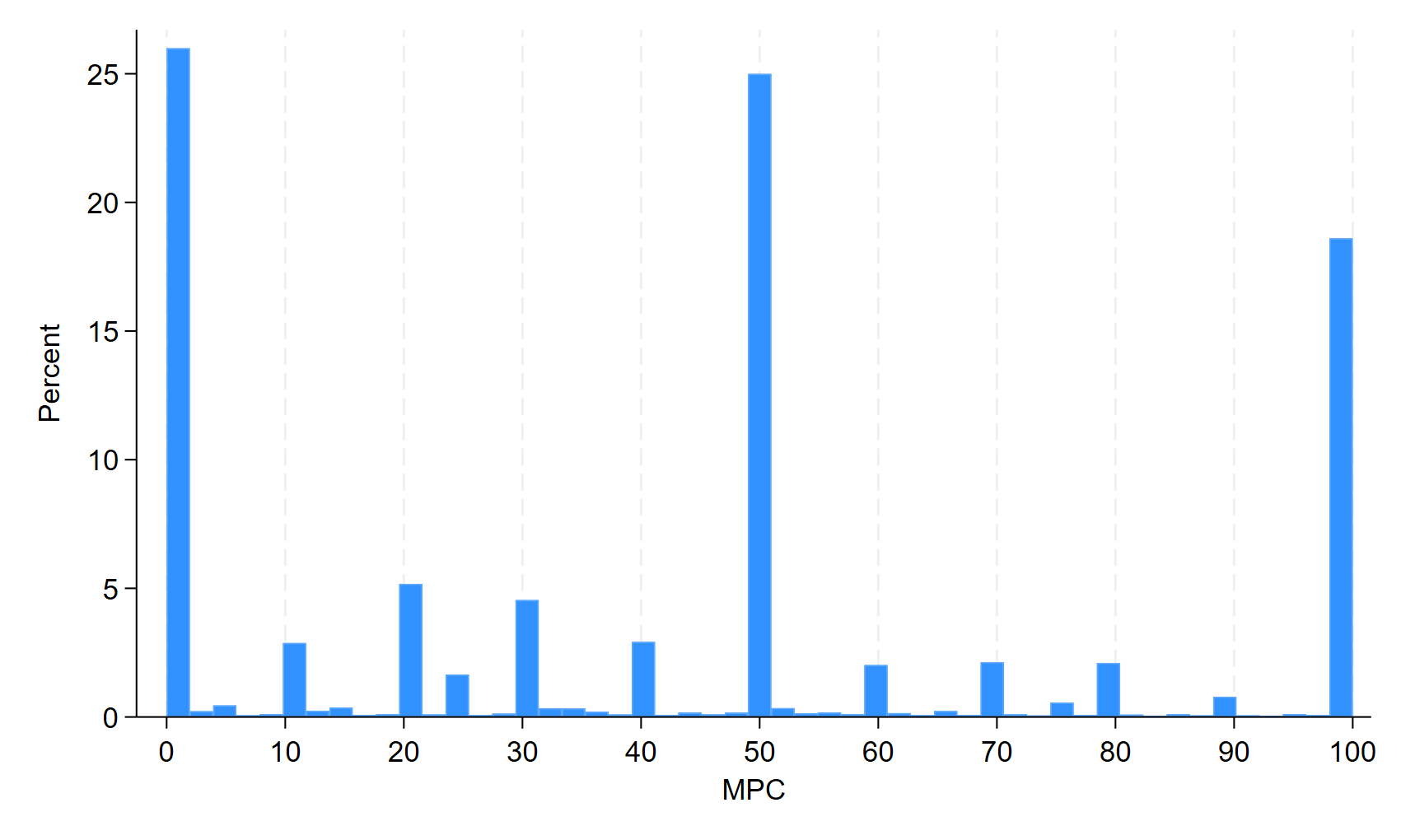}
    \begin{tablenotes}
        \footnotesize
        \item \textit{Notes:} The figure shows the distribution of the responses to the survey question eliciting the MPC. Sample size is 157,430. Data come from the Household Finance and Consumption Survey, 2017--2023. MPC in percent on horizontal axis. Response frequency in frequent on vertical axis.
    \end{tablenotes}
\end{threeparttable}

\end{figure}
\clearpage

\begin{figure}[H]
\centering

\caption{Histogram of the distribution on the change in MPC}
\label{Fig::MPC_change_distribution}

\begin{threeparttable}
    \includegraphics[width=1\textwidth]{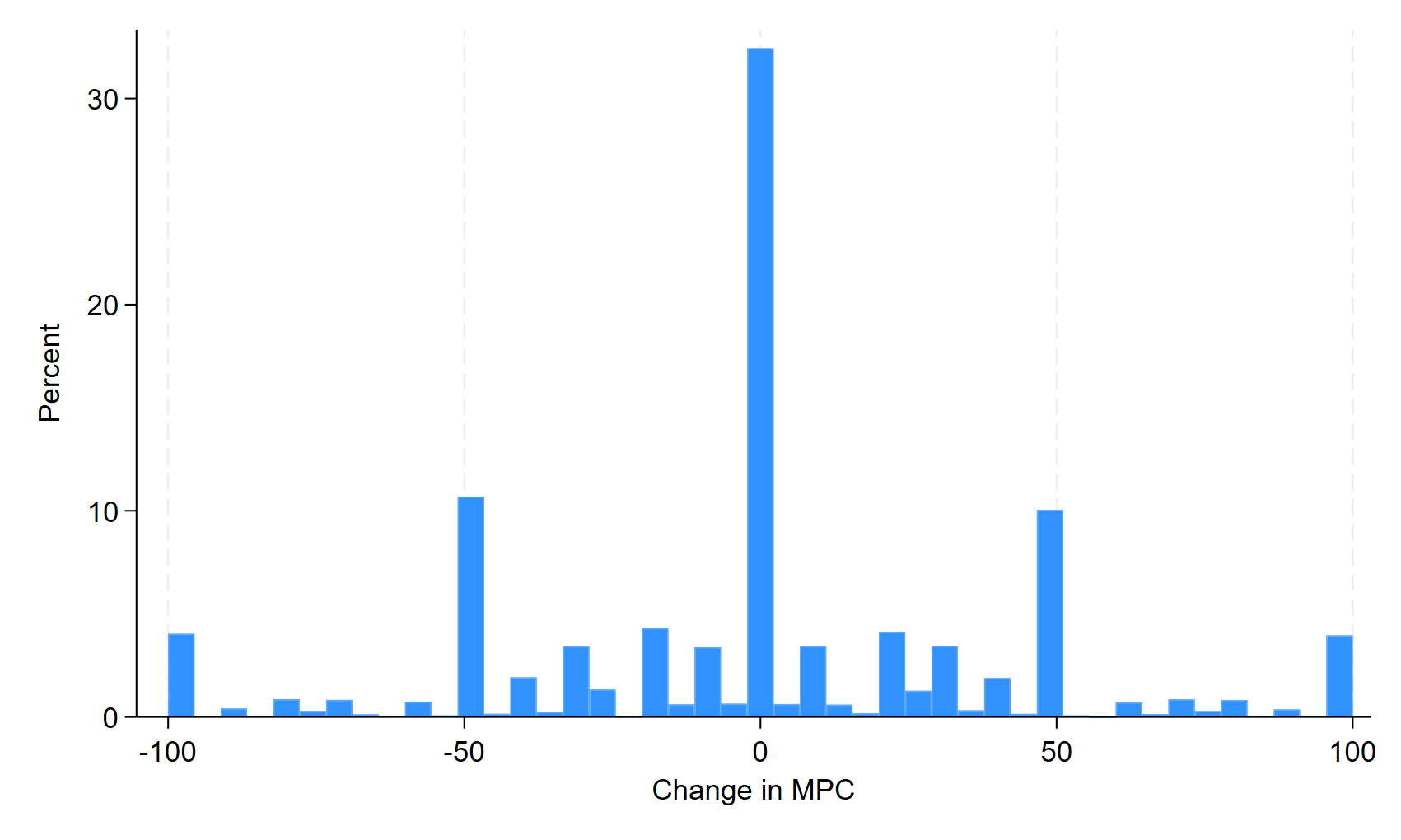}
    \begin{tablenotes}
        \footnotesize
        \item \textit{Notes:} The figure shows the distribution of the changes in reported MPCs. Sample size is 36,788. Data come from the Household Finance and Consumption Survey, 2017--2023. The change in MPC in percent on horizontal axis. Response frequency in frequent on vertical axis.
    \end{tablenotes}
\end{threeparttable}

\end{figure}
\clearpage

\begin{table}[H]
\begin{center}
\caption{Sample composition by country, unbalanced panel sample}\label{tab::Composition}
\begin{tabular}{lcc}
\toprule
Country & \# households & \# observations \\ \midrule
Belgium & 1,234 & 2,897 \\
Cyprus & 901 & 2,310 \\
Germany & 2,823 & 7,244 \\
Estonia & 1,243 & 2,486 \\
Spain & 4,621 & 11,052 \\
France & 7,669 & 18,342 \\
Italy & 4,385 & 10,616 \\
Lithuania & 303 & 644 \\
Latvia & 299 & 598 \\
Malta & 445 & 890 \\
Netherlands & 1,575 & 3,891 \\
Slovakia & 1,316 & 2,632 \\
\textbf{Pooled} & \textbf{26,814} & \textbf{63,602} \\
\bottomrule
\end{tabular} 
\begin{minipage}{0.7\textwidth}
\footnotesize
\textit{Notes}: Data come from the Household Finance and Consumption Survey, 2017--2023. Sample is restricted to households whose head is between 22 and 79 years old, have no negative family income and whose income is not entirely derived from self-employment, observed for at least two periods.
\end{minipage}
\end{center}
\end{table}
\clearpage

%%%%%%%%%%%%%%%%%%%%%%%%%%%%%%%%%%%%%%%%
%%% End document
%%%%%%%%%%%%%%%%%%%%%%%%%%%%%%%%%%%%%%%%

\end{document}